\documentclass[reprint,prl,aps]{revtex4-1}
\usepackage{graphicx}
\usepackage{color}
\usepackage{dcolumn}
\usepackage{bm}
\usepackage{amssymb}
\usepackage{color,amsmath}
\usepackage{soul,xcolor} 
\setstcolor{red} 
\usepackage{epstopdf}

\begin{document}

\title{Electrically tunable correlated and topological states in twisted monolayer-bilayer graphene}

\author{Shaowen Chen$^{1,2*}$}
\author{Minhao He$^{3*}$}
\author{Ya-Hui Zhang$^{4}$}
\author{Valerie Hsieh$^{1}$}
\author{Zaiyao Fei$^{3}$}
\author{K. Watanabe$^{5}$} 
\author{T. Taniguchi$^{5}$} 
\author{David H. Cobden$^{3}$}
\author{Xiaodong Xu$^{3,6\dagger}$}
\author{Cory R. Dean$^{1\dagger}$}
\author{Matthew Yankowitz$^{3,6\dagger}$}

\affiliation{$^{1}$Department of Physics, Columbia University, New York, NY, USA}
\affiliation{$^{2}$Department of Applied Physics and Applied Mathematics, Columbia University, New York, NY, USA}
\affiliation{$^{3}$Department of Physics, University of Washington, Seattle, WA, USA}
\affiliation{$^{4}$Department of Physics, Harvard University, Cambridge, MA, USA}
\affiliation{$^{5}$National Institute for Materials Science, 1-1 Namiki, Tsukuba 305-0044, Japan}
\affiliation{$^{6}$Department of Materials Science and Engineering, University of Washington, Seattle, WA, USA}
\affiliation{$^{*}$These authors contributed equally to this work.}
\affiliation{$^{\dagger}$xuxd@uw.edu (X.X.); cd2478@columbia.edu (C.R.D.); myank@uw.edu (M.Y.)}


\maketitle

\textbf{Twisted van der Waals heterostructures with flat electronic bands have recently emerged as a platform for realizing correlated and topological states with an extraordinary degree of control and tunability~\cite{Cao2018a,Cao2018b,Yankowitz2019,Lu2019,Stepanov2019,Saito2019,Sharpe2019,Serlin2020,Arora2020,Shen2020,Liu2019,Cao2019,Burg2019,He2020,Chen2019,Chen2019sctrilayer,Chen2020,Tang2020,Regan2020,Wang2019}. In graphene-based moir\'e heterostructures, the correlated phase diagram and band topology depend strongly on the number of graphene layers~\cite{Cao2018a,Cao2018b,Yankowitz2019,Sharpe2019,Lu2019,Serlin2020,Stepanov2019,Saito2019,Arora2020,Shen2020,Liu2019,Cao2019,Burg2019,He2020,Chen2019,Chen2019sctrilayer,Chen2020}, their relative stacking arrangement~\cite{Chen2019,Chen2019sctrilayer,Chen2020}, and details of the external environment from the encapsulating crystals~\cite{Chen2019,Chen2019sctrilayer,Chen2020,Sharpe2019,Serlin2020,Arora2020}. Here, we report that the system of twisted monolayer-bilayer graphene (tMBG) hosts a variety of correlated metallic and insulating states, as well as topological magnetic states. Because of its low symmetry, the phase diagram of tMBG approximates that of twisted bilayer graphene when an applied perpendicular electric field points from the bilayer towards the monolayer graphene, or twisted double bilayer graphene when the field is reversed. In the former case, we observe correlated states which undergo an orbitally driven insulating transition above a critical perpendicular magnetic field. In the latter case, we observe the emergence of electrically tunable ferromagnetism at one-quarter filling of the conduction band, with a large associated anomalous Hall effect. Uniquely, the magnetization direction can be switched purely with electrostatic doping at zero magnetic field. Our results establish tMBG as a highly tunable platform for investigating a wide array of tunable correlated and topological states.}

In select van der Waals (vdW) heterostructures, coupling to a moir\'e superlattice renormalizes the band structure and results in nearly flat electronic bands. These systems act as highly-tunable platforms capable of hosting a variety of correlated and topological ground states, including correlated insulating (CI) states, superconductivity, and ferromagnetism with an associated quantum anomalous Hall effect (QAHE)~\cite{Cao2018a,Cao2018b,Yankowitz2019,Lu2019,Stepanov2019,Saito2019,Sharpe2019,Serlin2020,Arora2020,Shen2020,Liu2019,Cao2019,Burg2019,He2020,Chen2019,Chen2019sctrilayer,Chen2020,Tang2020,Regan2020,Wang2019}. In moir\'e heterostructures composed entirely of graphene, the correlated phase diagram depends sensitively on the number of graphene sheets. In twisted bilayer graphene (tBLG) --- two rotated sheets of monolayer graphene --- CI states and superconductivity can be induced and tuned by changing the twist angle~\cite{Cao2018a,Cao2018b,Yankowitz2019,Lu2019,Stepanov2019,Saito2019} and by applying pressure~\cite{Yankowitz2019}. In addition, ferromagnetism and a QAHE can be obtained in the system by further rotationally aligning tBLG with a boron nitride (BN) substrate~\cite{Sharpe2019,Serlin2020}. In twisted double bilayer graphene (tDBG) --- two rotated sheets of Bernal-stacked bilayer graphene ---  spin-polarized CI states are observed that can be further tuned by application of a perpendicular electric field~\cite{Shen2020,Liu2019,Cao2019,Burg2019,He2020}. Finally, three layers of rhombohedrally-stacked graphene rotationally aligned with BN can also manifest CI states and ferromagnetism with non-trivial band topology~\cite{Chen2019,Chen2019sctrilayer,Chen2020}.   

\begin{figure*}[ht]
\includegraphics[width=6in]{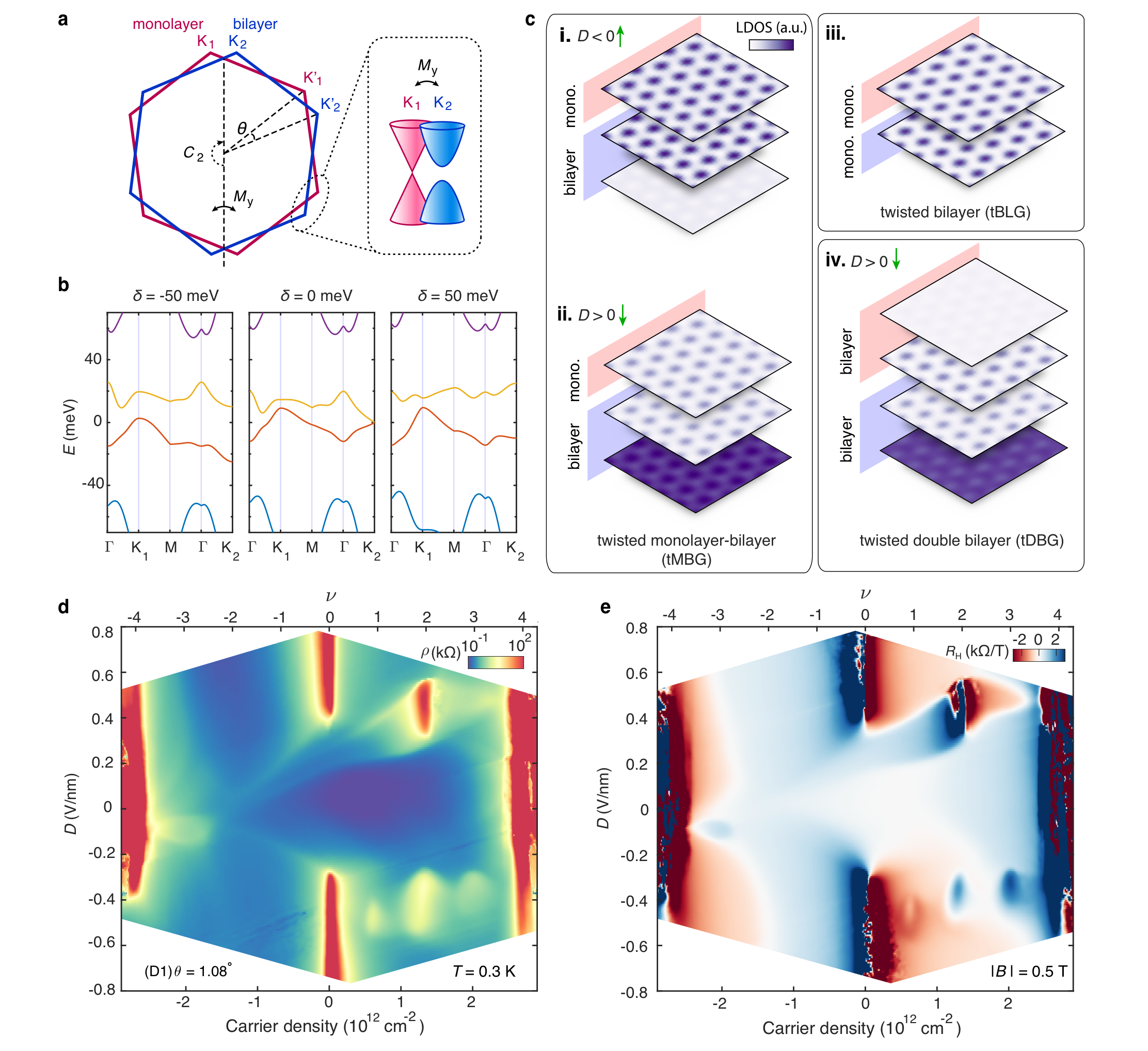} 
\centering
\caption{\textbf{Electronic structure of tMBG and transport in a device with $\theta = 1.08 ^{\circ}$.}
\textbf{a}, First Brillouin zones of monolayer (red) and bilayer (blue) graphene, rotated by a twist angle $\theta$. Both $C_2$ rotation symmetry and $M_y$ mirror symmetry are broken in tMBG.
\textbf{b}, Calculated band structure of tMBG with $\theta = 1.08 ^{\circ}$ at various values of interlayer potential, $\delta$.
\textbf{c}, Corresponding calculated local density of states for each layer of tMBG at full filling of the conduction band with (\textbf{i}) $D<0$ and (\textbf{ii}) $D>0$. Green arrows indicate the direction of the electric field. They are compared with the LDOS calculated for (\textbf{iii}) tBLG and (\textbf{iv}) tDBG at the same twist angle. The rotated interface is between layers with different color projections (red or blue).
\textbf{d}, Resistivity of device D1 at $T=0.3$~K. The corresponding band filling factor $\nu$ is shown on the top axis. 
\textbf{e}, Corresponding Hall coefficient, $R_\mathrm{H}$, antisymmetrized at $|B|=0.5$~T.
}
\label{fig:1}
\end{figure*}

Heterostructures comprising rotated sheets of monolayer and Bernal-stacked bilayer graphene have so far not been investigated experimentally, but are theoretically promising candidates to host correlated and topological states owing to their flat electronic bands with non-zero Chern numbers~\cite{SuarezMorell2013,Ma2019,Liu2019theory}. Twisted monolayer-bilayer graphene (tMBG) has low crystal symmetry, breaking both two-fold rotation, $C_2$, and mirror reflection, $M_y$. The former exchanges opposite valleys within the same layer, and the latter exchanges the top and bottom components of the twisted heterostructure (Fig.~\ref{fig:1}a). Consequentially, these states are anticipated to be uniquely tunable with a simple combination of a single twist angle, $\theta$, gate doping, $n$, and perpendicular displacement field, $D$, without any additional complicating conditions involving an internal crystal stacking configuration~\cite{Chen2019,Chen2019sctrilayer,Chen2020} or coupling to a secondary moir\'e pattern~\cite{Sharpe2019,Serlin2020}. 

Here, we investigate electrical transport in dual-gated tMBG devices over a small range of twist angles, $0.89^{\circ}<\theta<1.55^{\circ}$. We find that the correlated phase diagram of tMBG can approximately resemble either tBLG or tDBG depending on the direction of $D$. When $D$ points from the bilayer towards the monolayer graphene, we observe correlated states at all integer filling factors similar to tBLG, with the exception that superconductivity does not occur. Although correlations appear to be weaker than in tBLG, CI states emerge above a critical perpendicular magnetic field. When the direction of the electric field is flipped, we observe a spin-polarized CI state at half filling of the conduction band similar to tDBG. Remarkably, tMBG additionally manifests correlated magnetic states with intrinsic topological order in this configuration, which can be controlled by $\theta$, $D$, and $n$. Our work thus provides a means of realizing a wide array of tunable correlated and topological states within a single material platform.

\begin{figure*}[ht]
\includegraphics[width=6.9 in]{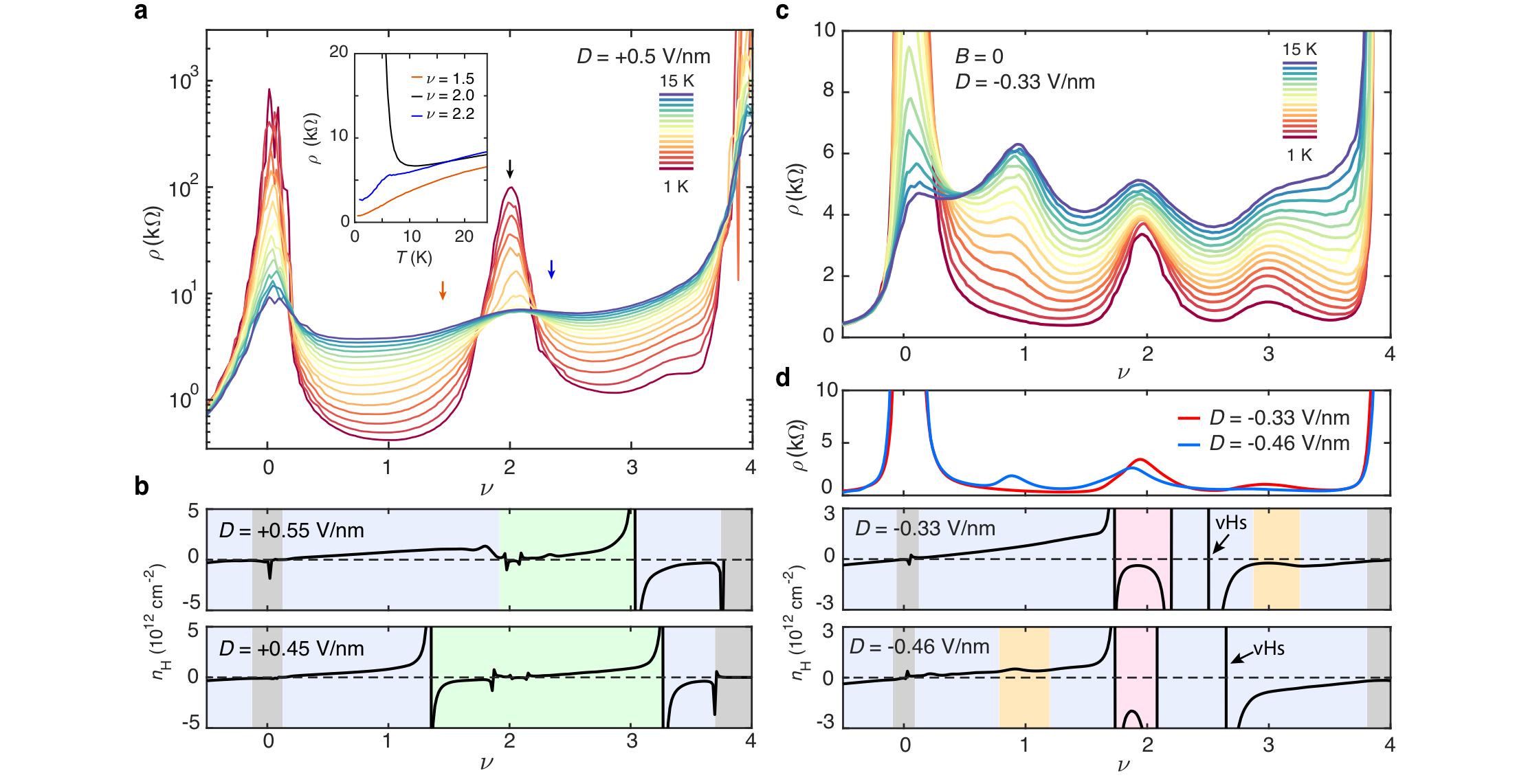} 
\caption{\textbf{Transport characteristics of correlated states in tMBG.}
\textbf{a}, Temperature dependence of $\rho$ at $D=+0.5$~V/nm. (Inset) $\rho(T)$ acquired at the filling factors denoted by the colored arrows in the main panel.
\textbf{b}, Hall density, $n_\mathrm{H}$, as a function of $\nu$ at various $D$. Color shading denotes regions of the phase diagram with different inferred degeneracy, where blue corresponds to four-fold degeneracy and green corresponds to two-fold degeneracy. Gray-shaded regions denote insulating states at $\nu=0$ and 4.
\textbf{c}, Temperature dependence of $\rho$ at $D=-0.33$~V/nm.
\textbf{d}, (Top panel) $\rho$ vs. $\nu$ at selected $D$ and $T=0.3$~K. (Bottom panels) Corresponding $n_\mathrm{H}$. Color shading is as in \textbf{b}, but with pink denoting regions with correlation-driven sign changes in $n_\mathrm{H}$, and gold denoting regions with deviations from the anticipated gate-induced density $n$. An additional sign change related to the single-particle vHs is denoted by an arrow.
}
\label{fig:2}
\end{figure*}

Figure~\ref{fig:1}b shows the continuum model calculation of the band structure of tMBG at a twist angle of $\theta = 1.08^{\circ}$ as a function of the interlayer potential difference, $\delta$ (see Supplementary Information). In contrast to tBLG and tDBG, because of the low symmetry the band structure of tMBG differs depending on the direction of $D$ (i.e. the sign of $\delta$). Additionally, calculations of the layer-resolved local density of states (LDOS) reveal that it is favorable for charge to be arranged highly asymmetrically amongst the three constituent graphene layers. Panels \textbf{i} and \textbf{ii} of Fig.~\ref{fig:1}c show the LDOS of tMBG calculated at full filling of the conduction band for opposite signs of $D$. In both cases, the moir\'e potential localizes the LDOS on a triangular lattice of AAB-stacked sites (analogous to AA-stacked sites in tBLG, in which the carbon atoms sit directly atop one another). For $D<0$ ($\delta<0$), when the field points from the bilayer towards the monolayer graphene, the LDOS on the outer sheet of the bilayer graphene is suppressed, and the LDOS on the two graphene layers at the rotated interface resembles that of tBLG (panel \textbf{iii}). On the other hand, for $D>0$ ($\delta>0$), the LDOS on the outer graphene layer is instead enhanced. In this case the LDOS resembles that of tDBG (panel \textbf{iv}), where the fourth (uppermost) graphene layer has a small LDOS. We would therefore anticipate that tMBG will act similarly to either tBLG or tDBG depending on the sign of $D$, assuming the additional or absent layer of lightly doped graphene plays a limited role.

This picture is borne out well by the experiments. Figure~\ref{fig:1}d shows the resistivity, $\rho$, of a tMBG device with $\theta =1.08^{\circ}$ (device D1) as a function of carrier density, $n$, and $D$ at temperature $T=0.3$~K. The top axis indicates the filling factor of the bands, $\nu$ (see Methods). Qualitatively, many of the observed features can be associated with the band structure calculated in Fig.~\ref{fig:1}b (also Supplementary Information Fig.~\ref{fig:S_calculated_energy}). In particular, at $\nu=0$ the device behaves as a semi-metal for small $|D|$, but becomes semiconducting at larger $|D|$. Insulating features are observed over a wide range of $|D|$ at full filling of the conduction or valence band ($\nu=\pm4$). Additional resistance peaks disperse as a function of $D$ within the flat bands, and likely correspond to van Hove singularities (vHs) as in tDBG~\cite{He2020}.

In the conduction band ($0<\nu<+4$), additional resistive peaks are observed over a finite range of $D$ that appear to have no counterparts in the single-particle model. These resistive peaks are observed at integer $\nu$, each within a different finite range of $D$. Notably, these features are inequivalent for opposite signs of $D$, in contrast with tBLG and tDBG where the correlated states are approximately symmetric with $D$. Measurements of the antisymmetrized Hall coefficient, $R_\mathrm{H} = (R_{xy}[B] - R_{xy}[-B])/(2B)$, provide additional insight into their nature (Fig.~\ref{fig:1}e). For $D>0$, there is a highly resistive state at half filling surrounded by a stretched halo of somewhat increased resistance. As in tDBG, $R_\mathrm{H}$ changes sign within this extended halo region, suggestive of a spontaneously broken symmetry within the bands~\cite{He2020}. In contrast, for $D<0$ there are three weakly resistive features at each integer $\nu$, and abrupt changes in $R_\mathrm{H}$ occur only near these features. 

\begin{figure*}[ht]
\includegraphics[width=6.9 in]{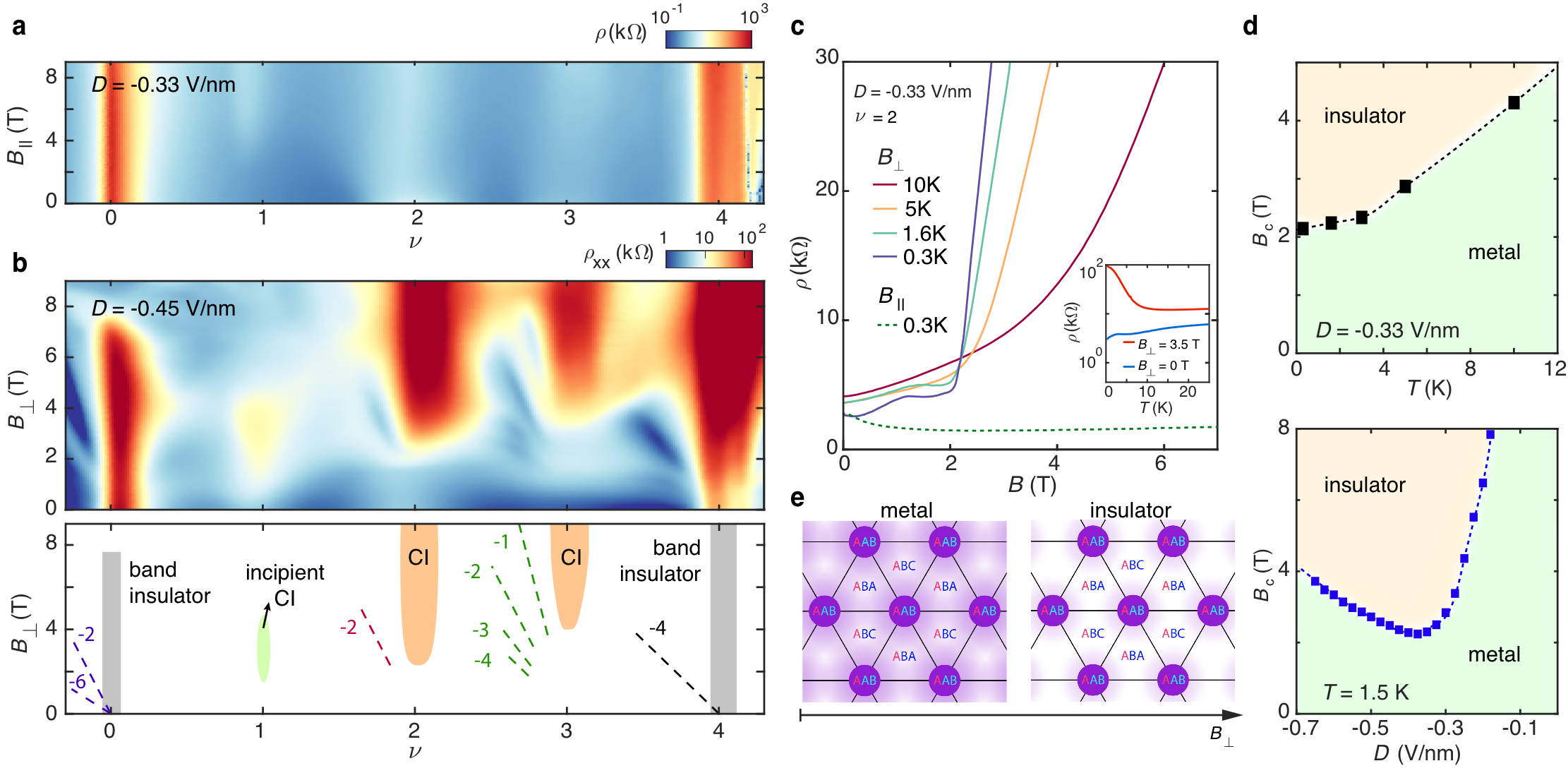} 
\caption{\textbf{Properties of the correlated states at $\nu=1,2,3$ for $D<0$.}
\textbf{a}, Device resistivity up to full filling of the moir\'e unit cell with $D=-0.33$~V/nm as a function of $B_{||}$ at $T=300$~mK.
\textbf{b}, Landau fan diagram with $D=-0.45$~V/nm. The bottom panel schematically denotes the observed quantum oscillations, as well as band insulator states (gray), CI states (orange), and an incipient CI state (green). CI states at $\nu=2,3$ and associated quantum oscillations only emerge for finite $B_\perp$.
\textbf{c} $\rho(B_\perp)$ at various $T$ (solid curves) and $\rho(B_{||})$ at $T=300$~mK. (Inset) $\rho(T)$ at $\nu=2$ for $B_\perp=0$ and 3.5~T. 
\textbf{d}, Critical magnetic field for the onset of insulating behavior, $B_c$, versus $T$ (top) and $D$ (bottom). Dashed lines are guides to the eye. Regions colored green correspond to metallic temperature dependence, whereas beige regions correspond to insulating behavior.
\textbf{e}, Cartoon schematics of the inferred LDOS (light pink) on the moir\'e lattice (AAB sites denoted by dark purple dots) as a function of $B_\perp$.
}
\label{fig:3}
\end{figure*}

Figure~\ref{fig:2}a shows the temperature dependence of $\rho$ in the conduction band at $D=+0.5$~V/nm, in which we expect the LDOS distribution to resemble that of tDBG. We observe clear insulating behavior at half filling ($\nu=2$), with energy gaps that appear to grow upon applying an in-plane magnetic field indicative of spin-polarized ordering (Supplementary Information Fig.~\ref{fig:S_parallelB}). Slightly away from half filling, we observe abrupt drops in $\rho$ as the temperature is lowered (blue curve in the inset of Fig.~\ref{fig:2}a). These abrupt drops arise within regions of the phase diagram in which the magnitude or sign of the Hall density, $n_\mathrm{H} \equiv e/R_\mathrm{H}$, departs from single-particle expectations (green shaded regions in Fig.~\ref{fig:2}b). These features correspond closely with the CI and metallic states in tDBG~\cite{Shen2020,Liu2019,Cao2019,Burg2019,He2020}, suggesting that at this twist angle the phase diagram of tMBG is similar to that of tDBG, with small differences arising owing to the absence of the weakly charged fourth graphene sheet (see Fig.~\ref{fig:1}c, \textbf{iv}).

Figure~\ref{fig:2}c shows similar temperature dependence of $\rho$ at $D=-0.33$~V/nm, in which we expect the LDOS distribution to resemble that of tBLG. The states at $\nu=1,2,3$ all exhibit metallic temperature dependence, defined by a reduction in $\rho$ as $T$ is lowered, in direct contrast to the behavior of the CI state for $D>0$. Notably, the feature at $\nu=1$ is apparent at high temperature ($T=15$~K in Fig.~\ref{fig:2}c), but becomes suppressed at low temperature. Qualitatively similar behavior has been observed previously in tBLG, most clearly in devices with twist angles slightly detuned from the ``magic angle''~\cite{Yankowitz2019} and in devices with strong electrostatic screening from a nearby metal gate~\cite{Stepanov2019}. This suggests that our devices are marginally correlated compared to magic angle tBLG, potentially owing to a combination of larger bandwidth (Supplementary Information Fig.~\ref{fig:S_calculated_energy}) and electrostatic screening from the addition of the weakly charged third graphene sheet. However, unlike in tBLG, these states can be tuned with $D$. For example, Fig.~\ref{fig:2}d shows $\rho$ for two values of $D$, in which the state at $\nu=1$ is apparent at $D=-0.46$~V/nm but not at $D=-0.33$~V/nm. These correlated metallic states are also associated with anomalies in $n_\mathrm{H}$ (Fig.~\ref{fig:2}d); either a sign change (pink shading) or deviations from a linear gate-induced charge density, $n$ (gold shading). An additional sign change in $n_\mathrm{H}$ is observed for $2<\nu<3$ that matches the vHs in the single-particle band structure (see also Supplementary Information Fig.~\ref{fig:S_HallDen}), suggesting that correlations do not greatly modify the conduction band Fermi surface at filling factors away from the integers.

\begin{figure*}[ht]
\includegraphics[width=5in]{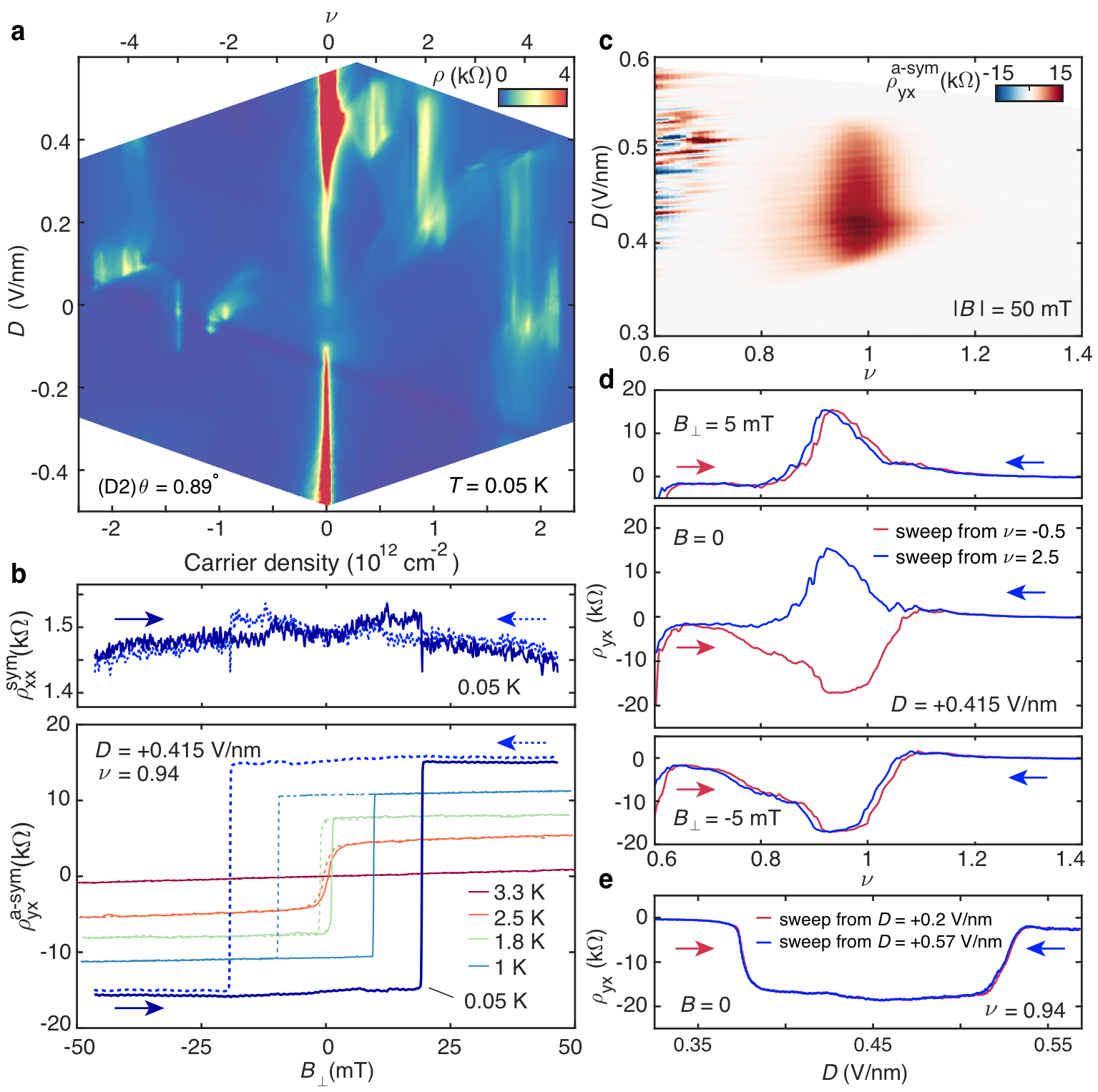} 
\caption{\textbf{Ferromagnetism and AHE at $\nu=1$ for $D>0$ in a device with $\theta = 0.89^{\circ}$.}
\textbf{a}, Resistivity of device D2 at $T=0.05$~K. The corresponding band filling factor $\nu$ is shown on the top axis. 
\textbf{b}, Symmetrized $\rho_{xx}$ and antisymmetrized $\rho_{yx}$ acquired as $B_\perp$ is swept back and forth at $\nu=0.94$ and $D=+0.415$~V/nm with $T=0.05$~K (blue curves). $\rho_{yx}$ is additionally shown at different temperatures. Arrows indicate the direction of the magnetic field sweep, corresponding to solid and dashed curves.
\textbf{c}, $\rho_{yx}$, antisymmetrized at $|B|=50$~mT. Rapidly oscillating red and blue points near $\nu \approx 0.6$ are related to the nearby insulating charge neutrality point rather than magnetic order.
\textbf{d}, $\rho_{yx}$ as $\nu$ is swept back and forth from $-0.5$ to $+2.5$, acquired at $B_\perp=5$~mT, 0, and $-5$~mT (top to bottom). $B=0$ corresponds to zero current in the superconducting magnet coil, with the offset due to trapped flux being less than $\sim$1~mT.
\textbf{e}, $\rho_{yx}$ as $D$ is swept back and forth from $+0.2$~V/nm to $+0.57$~V/nm at $B=0$.
}
\label{fig:4}
\end{figure*}

The correlated metallic states observed for $D<0$ have little to no dependence on magnetic field applied in-plane, $B_{||}$ (shown up to 9 T in Fig.~\ref{fig:3}a). However, applying a perpendicular field, $B_\perp$, causes a sharp transition to an insulating behavior at $\nu=2$ and 3 above a critical field value (Fig.~\ref{fig:3}b). Signs of an incipient resistive state can also be seen over a finite range of $B_\perp$ at $\nu=1$. Quantum oscillations associated with the states at $\nu=2$ and 3 emerge only above a finite $B_\perp$, with apparent two-fold degeneracy neighboring $\nu=2$ and no remaining degeneracies neighboring $\nu=3$ (Fig.~\ref{fig:3}b, lower panel). This too is consistent with prior studies of tBLG~\cite{Yankowitz2019,Saito2019,Stepanov2019}, in which recent measurements have indicated that a Dirac-like dispersion is revived at each quarter band filling owing to a cascade of spontaneously broken symmetries~\cite{Zondiner2019,Wong2019}.

Figure~\ref{fig:3}c shows $\rho(B_\perp)$ at $\nu=2$ for various temperatures (solid curves), as well as $\rho(B_{||})$ at $T=0.3$~K (dashed curve). While $\rho$ is nearly insensitive to $B_{||}$, we observe an abrupt transition to insulating behavior above a critical perpendicular magnetic field, $B_{c}$, at each temperature, defined as the crossover point between separate linear fits to $\rho(B_{||})$ at low and high $B_\perp$. The inset of Figure~\ref{fig:3}c compares $\rho(T)$ at $B_\perp=0$ and 3.5~T for $\nu=2$, showing the metallic and insulating behavior, respectively (see Supplementary Information Fig.~\ref{fig:S_Bperp_all} for $\nu=1,3$). We find that $B_{c}$ first grows weakly with increasing temperature, then increases rapidly at higher temperature (Fig.~\ref{fig:3}d). $B_c$ can also be tuned with $D$, as shown for $T=1.5$~K in (Fig.~\ref{fig:3}d). It is peculiar that insulating behavior onsets with $B_\perp$ but has almost no dependence on $B_{||}$. In the simple approximation of a perfectly 2D sample, $B_\perp$ couples to both the spin and orbital Zeeman energies whereas $B_{||}$ only couples to the spin Zeeman energy. Consequentially, spin-polarized (unpolarized) states are typically strengthened (suppressed) by a magnetic field pointing in either orientation. The correlated states for $D<0$ do not appear to have a natural interpretation in this context. 

Insulating states emerging with $B_\perp$ have been previously observed in bulk correlated systems including Bechgaard salts~\cite{Vignolles2005} and high-mobility semiconductors such as Hg$_{1-x}$Cd$_{x}$Te~\cite{Rosenbaum1985,Field1988}. These insulating states were thought to be driven by orbital confinement effects, associated with spin-density waves in the former and Wigner crystal formation in the latter. The overall similarity of our $B_c(T)$ to those results suggests that related orbital confinement effects may be at play in tMBG. The cartoon schematic in Fig.~\ref{fig:3}e qualitatively illustrates the presumed atomic-scale electronic structure of tMBG in the metallic and field-assisted insulating phases. The dark purple sites correspond to regions of AAB stacking, and the the lighter pink halo represents the associated LDOS. $B_\perp$ is expected to orbitally confine the LDOS more tightly around the AAB sites, resulting in an enhancement in the effective interaction strength. In contrast, the LDOS does not depend strongly on $B_{||}$ as orbital effects are anticipated to be weak owing to the atomically-thin nature of tMBG, and therefore the interaction strength is also not substantially modified.

Qualitatively similar insulating states with associated quantum oscillations have previously been observed to arise abruptly at finite $B_\perp$ in tBLG devices slightly detuned from the magic angle with strong electrostatic screening from a nearby gate~\cite{Stepanov2019}. However, the behavior of the states at integer filling with $B_{||}$ was not reported, therefore a full comparison is not possible. Nevertheless, the overall similarity of these results supports our model of tMBG with $D<0$ as approximately analogous to electrostatically screened tBLG, in which the correlation strength is marginal. One important differnce, however, is that superconductivity appears to be absent in our device, whereas it is a robust ground state in screened tBLG~\cite{Stepanov2019,Saito2019,Liu2020}. Differences in bandwidth may be a contributing factor (see Supplementary Information Fig.~\ref{fig:S_calculated_energy}), however it is also possible that the low symmetry of tMBG plays a role. Superconductivity may be sensitive to the combination of $C_2$~\cite{Khalaf2020} and $M_y$ symmetries, both broken in tMBG. So far there have been no unambiguous report of superconductivity in a moir\'e vdW platform with broken $C_2$ symmetry~\cite{He2020}.

Finally, we turn to the correlated states emerging for $D>0$ in a second device (device D2) which has a slightly smaller twist angle, $\theta =0.89^{\circ}$. Figure~\ref{fig:4}a shows the resistivity of device D2 at $T=0.05$~K and $B=0$~T. The overall behavior is similar to device D1, with a few key exceptions. First, the insulating behavior is weak or absent at $\nu=\pm4$, indicating the energy gaps to the remote bands are small or zero at this twist angle. Second, incipient CI states can be seen at $\nu=-2$ and $-3$ for $D \approx 0$ (Supplementary Information Fig.~\ref{fig:S_ahe_RvT}), and have no known analog in tDBG. Third, an additional resistive state emerges at $\nu=1$ over a small range of $D>0$. Remarkably, this state exhibits a robust anomalous Hall effect (AHE) with large Hall angle $\rho_{yx}/\rho_{xx} \approx 10$ at $B=0$ (blue curves in Fig.~\ref{fig:4}b). $\rho_{yx}$ exhibits hysteresis about zero magnetic field, an unambiguous characteristic of ferromagnetism. A single loop with a coercive field of $B_\perp \approx20$~mT is formed as the field is swept back and forth. We field symmetrize $\rho_{xx}$ and antisymmetrize $\rho_{yx}$ to eliminate small offsets in the two owing to mixing (see Supplementary Information Fig. \ref{fig:S_ahe_raw}). The temperature dependence of the ferromagnetism is additionally shown in Fig.~\ref{fig:4}b. The hysteresis and magnitude of the AHE are suppressed upon raising the temperature, and vanish above $T\approx 2.5$~K (Supplementary Information Fig.~\ref{fig:S_ahe_Tdep}). The sharpness of the magnetization flip implies that the easy axis is perpendicular to the sample, and that a single magnetic domain is dominant within the measured region of the sample. 

Figure~\ref{fig:4}c shows a map of $\rho_{yx}$ antisymmetrized at $|B|=50$~mT. The regime of ferromagnetic order as a function of $\nu$ and $D$ is indicated by the dark red at $\nu \approx 1$. The ferromagnetism occurs over an extended range of $D$, and persists into the metallic states slightly outside of the $\nu=1$ gap. We do not observe any signs of ferromagnetic ordering at any other filling factors in this device. Despite the fact that $\rho_{yx}$ is not quantized to an integer fraction of $h/e^2$, and $\rho_{xx}$ does not vanish, the large Hall angle indicates transport through edge states. Nonlocal resistance measurements provide further evidence supporting chiral edge conduction in our sample (see Supplementary Information Fig. \ref{fig:S_ahe_NL})~\cite{Kou2014,Sharpe2019}. It seems likely that a fully developed Chern insulator state could be realized in a sample with lower disorder or a larger energy gap. In a model of the isolated conduction band of tMBG as four overlapping flat bands with varying spin and valley order, spontaneous polarization of carriers into a single symmetry-broken band with non-zero Chern number, and consequent QAHE, has been predicted~\cite{Sharpe2019,Serlin2020,Repellin2019,Bultinck2019,Zhang2019ahe}. However, more exotic ground states with mixed spin and valley polarization may be favored, and our experiment is not sensitive to the exact ground state polarization.

In addition to switching the magnetic ordering with $B_\perp$, we are able to control the magnetic state purely with charge doping at $B=0$. Figure~\ref{fig:4}d shows $\rho_{yx}$ as the doping is swept back and forth, exhibiting opposite signs of the magnetic ordering for each sweep direction. In contrast, we do not observe switching upon sweeping $D$ back and forth (Figure~\ref{fig:4}e). Doping-controlled switching of the magnetic order has been recently predicted owing to the unique orbital magnetism exhibited in moir\'e vdW heterostructures~\cite{Zhu2020}. However, in contrast to this prediction our switching arises only within a few millitelsa of $B=0$ (Supplementary Information Fig.~\ref{fig:S_ahe_InitiWithB}), and therefore likely originates from a different mechanism. Notably, the magnetic order does not switch precisely upon doping across $\nu=1$, but rather requires sweeping the doping far away from the magnetically-ordered regime before returning. The switching occasionally occurs through metastable intermediate states and is observed more reliably when $n$ is tuned slowly (Supplementary Information Fig.~\ref{fig:S_ahe_stable}), suggesting that it is an intrinsic effect. Furthermore, the switching does not appear to depend on any other initialization conditions (Supplementary Information Fig.~\ref{fig:S_ahe_fieldscans}). Although we are not able to identify the valley and spin ordering of the surrounding symmetry-broken bands at $\nu=1$, this effect appears to be tied to a selective polarization of each of these bands with doping. The exact mechanism responsible for these unusual switching dynamics --- including whether it is of fundamental origin or related to sample inhomogeneity --- remains an open question for future work.

The ability to electrically switch magnetic order is rare amongst known materials, and can typically be achieved with electric fields only in multiferroics~\cite{Matsukura2015}. tMBG therefore provides a new platform for spintronics applications with purely doping-controlled magnetism and ultra-low power dissipation. More generally, our results demonstrate that the low symmetry of tMBG enables a wide array of highly tunable correlated states which differ upon changing the sign of $D$. The addition of intrinsic non-trivial band topology also raises the possibility of realizing fractional Chern insulator states~\cite{Zhang2019} in samples with stronger correlations at different twist angle or under pressure~\cite{Yankowitz2019}. We anticipate that tMBG will be a crucial platform for further understanding of these correlated and topological states of matter, providing unprecedented control with external tuning parameters.

\section*{Methods}

tMBG devices are fabricated using the ``cut-and-stack'' method~\cite{Saito2019}. We identify exfoliated graphene flakes with regions of both monolayer and bilayer graphene, and use an atomic force microscope (AFM) tip to isolate a region from each area~\cite{Chen2019,Saito2019}. We find that in general, longer AFM-cut edges can increase the friction between rotated graphene sheets and improve the yield of devices near the targeted twist angle. Samples are assembled using a standard dry transfer technique that utilizes a polycarbonate (PC) film on top of a polydimethyl siloxane (PDMS) dome~\cite{wang_one-dimensional_2013}. Completed heterostructures are transferred onto a Si/SiO$_2$ wafer. The temperature is kept below 180$^\circ$C during device fabrication to preserve the intended twist angle. 

All devices are encapsulated between flakes of BN with typical thickness of $30-40$~nm, and have graphite top and bottom gates (except the device with twist angle of 1.44$^\circ$, which has a silicon bottom gate). In devices with a graphite bottom gate, a gate voltage is applied to the Si gate in order to dope the region of the graphene contacts overhanging the graphite back gate to a high charge carrier density and reduce the contact resistance. $n$ and $D$ can be tuned by the gate voltages through the relation $n = (C_\mathrm{TG}V_\mathrm{TG} + C_\mathrm{BG}V_\mathrm{BG})/e$, $D = |(C_\mathrm{TG}V_\mathrm{TG} - C_\mathrm{BG}V_\mathrm{BG})/2|$, where $C_\mathrm{TG}$ ($C_\mathrm{BG}$) is the capacitance between the top (bottom) gate and tMBG, and $e$ is the electron charge. We choose a convention such that $D>0$ corresponds to the field pointing from from monolayer to bilayer graphene.
 
We measure electrical transport in four tMBG devices with twist angles varying between $\theta=0.89^\circ$ and 1.55$^\circ$ (Supplementary Information Figs.~\ref{fig:S_devices},~\ref{fig:S_largerangle}, and~\ref{fig:S_1p44_Dall}). Transport measurements are conducted in a four-terminal geometry with ac current excitation of $10-20$~nA using standard lock-in techniques at either 13.3 Hz or 17.7 Hz. The magnetic hysteresis loops in device D2 are measured using a 500~pA excitation. A dc current bias, $I_{dc}$, can additionally be added to further control the magnetic state (Supplementary Information Fig.~\ref{fig:S_ahe_dcswitch}), as has been discussed previously in devices of tBLG aligned with BN~\cite{Sharpe2019,Serlin2020}. 
 
The twist angle $\theta$ is determined from the values of charge carrier density at which the insulating states at $\nu=\pm4 = \pm n_s $ are observed, following $n_s = 8\theta^2/\sqrt{3}a^2$, where $a=0.246$~nm is the lattice constant of graphene. The values of $\pm n_s$ are determined from the sequence of quantum oscillations in a magnetic field which project to $\pm n_s$ (or $\pm n_s/2$ for $\theta = 1.08^\circ$, see Supplementary Information Fig.~\ref{fig:DposQOs}.).

\section*{acknowledgments}
We thank A. Young, H. Polshyn, A. Millis, C.-Z. Chang, J.-H. Chu, and E. Khalaf for helpful discussions. Research on correlated states in twisted monolayer-bilayer graphene was primarily supported as part of Programmable Quantum Materials, an Energy Frontier Research Center funded by the U.S. Department of Energy (DOE), Office of Science, Basic Energy Sciences (BES), under award DE-SC0019443. Measurements and understanding of ferromagnetism at the University of Washington were partially supported by NSF MRSEC 1719797. X.X. acknowledges support from the Boeing Distinguished Professorship in Physics. X.X. and M.Y. acknowledge support from the State of Washington funded Clean Energy Institute. This work made use of a dilution refrigerator system which was provided by NSF DMR-1725221. K.W. and T.T. acknowledge support from the Elemental Strategy Initiative conducted by the MEXT, Japan and the CREST (JPMJCR15F3), JST.

\section*{Author contributions}
S.C., M.H. and V.H. fabricated the devices and performed the measurements. Y.-H. Z. performed the calculations. Z.F. and D.H.C. assisted with measurements in the dilution refrigerator. K.W. and T.T. grew the BN crystals. S.C., M.H., X.X., C.R.D. and M.Y. analyzed the data and wrote the paper with input from all authors.

\section*{Competing interests}
The authors declare no competing interests.

\section*{Data availability}
The data that support the plots within this paper and other findings of this study are available from the corresponding author upon reasonable request.

\bibliographystyle{naturemag}
\bibliography{references_twisted}

\clearpage

\renewcommand{\thefigure}{S\arabic{figure}}
\renewcommand{\thesection}{S\arabic{section}}
\renewcommand{\thesubsection}{S\arabic{subsection}}
\renewcommand{\theequation}{S\arabic{equation}}
\renewcommand{\thetable}{S\arabic{table}}
\setcounter{figure}{0} 
\setcounter{equation}{0}
\appendix 

\onecolumngrid

\section*{Supplementary Information}

\subsection{Calculation of Band Structure}\label{SI_calculation}

We calculate the band structure using the standard continuum model. The Hamiltonian is

\begin{equation}
	H=H_{MG}+H_{BG}+H_M
\end{equation}

We have  
\begin{equation}
	H_{MG}=\sum_{\mathbf k}(\tilde c^\dagger_{A}(\mathbf k), \tilde c^\dagger_{B}(\mathbf k))\left(
	\begin{array}{cc}
	M-\frac{\delta}{2}& -\frac{\sqrt{3}}{2}t(\tilde k_x-i \tilde k_y)\\
	-\frac{\sqrt{3}}{2}t(\tilde k_x+i \tilde k_y) &-M-\frac{\delta}{2} 
	\end{array}\right)  \left(\begin{array}{c} \tilde c_A(\mathbf k)\\ \tilde c_B(\mathbf k)\end{array}\right)
\end{equation}
where $\mathbf{\tilde k}=R(-\theta/2) \mathbf {k}$ with $\theta$ as the twist angle. $R(\varphi)$ is the transformation matrix for anticlockwise rotation with angle $\varphi$.

The Hamiltonian for the bilayer graphene is
\begin{equation}
	H_{BG}=\sum_{\mathbf k}\Psi^\dagger(\mathbf k)\left(
	\begin{array}{cccc}
	0& -\frac{\sqrt{3}}{2}t(\tilde k_x-i\tilde k_y) & -\frac{\sqrt{3}}{2}\gamma_4 (\tilde k_x-i \tilde k_y) & -\frac{\sqrt{3}}{2}\gamma_3(\tilde k_x+i\tilde k_y)\\
	-\frac{\sqrt{3}}{2}t(\tilde k_x+i\tilde k_y) &0 & \gamma_1  & -\frac{\sqrt{3}}{2}\gamma_4 (\tilde k_x-i \tilde k_y)\\
	-\frac{\sqrt{3}}{2}\gamma_4(\tilde k_x+i\tilde k_y) & \gamma_1 & -\frac{\delta}{2}& -\frac{\sqrt{3}}{2}t (\tilde k_x-i\tilde k_y)\\
	-\frac{\sqrt{3}}{2}\gamma_3(\tilde k_x-i\tilde k_y) &-\frac{\sqrt{3}}{2}\gamma_4(\tilde k_x+i\tilde k_y) & -\frac{\sqrt{3}}{2}t(\tilde k_x+i\tilde k_y) & \frac{\delta}{2}
	\end{array}\right)  \Psi(\mathbf k)
\end{equation}
where $\mathbf{\tilde k}=R(\theta/2)\mathbf{k}$ and $\Psi^\dagger(\mathbf k)=(c^\dagger_{A_1}(\mathbf k), c^\dagger_{B_1}(\mathbf k), c^\dagger_{A_2}(\mathbf k), c^\dagger_{B_2}(\mathbf k))$.

Finally the inter-layer moir\'e tunneling term is
\begin{equation}
	H_M=\sum_{\mathbf k} \sum_{j=0,1,2} (\tilde c^\dagger_{A}(\mathbf k), \tilde c^\dagger_{B}(\mathbf k))\left(
	\begin{array}{cc}
	\alpha t_M& t_M e^{-i  \frac{2\pi}{3} j}\\
	t_M e^{i  \frac{2\pi}{3} j} &\alpha t_M 
	\end{array}\right)  \left(\begin{array}{c} c_{A_1}(\mathbf k+\mathbf{Q}_j)\\ c_{B_1}(\mathbf k+\mathbf{Q}_j)\end{array}\right)+h.c.
\end{equation}
where $\mathbf Q_0=(0,0)$, $\mathbf Q_1=\frac{1}{a_M}(-\frac{2\pi}{\sqrt{3}},-2\pi)$ and $\mathbf Q_2=\frac{1}{a_M}(\frac{2\pi}{\sqrt{3}},-2\pi)$, with $a_M$ as the moir\'e lattice constant.

Here $\tilde c_\alpha(\mathbf k)$ and $c_\alpha(\mathbf k)$ are electron operators for the monolayer graphene and the bilayer graphene respectively.  We use parameters $(t, \gamma_1, \gamma_3, \gamma_4)=(-2610, 361, 283, 140)$ meV. For the inter-layer tunning, we use $t_M=110$ meV and $\alpha=0.5$.  $M$ is a sublattice potential term for the monolayer graphene which may originate from hBN alignment. We use $M=10$ meV. $\delta$ is the potential difference which is tuned by the displacement field $D$.

With sufficient $|\delta|$, the neutrality gap is opened and we can define valley Chern number for the conduction and valence bands. Here we focus on the conduction band, the valley Chern number is $|C|=2$ and $|C|=1$ for $D>0$ and $D<0$ sides respectively.
\clearpage

\begin{figure*}[t]
\centering
\includegraphics[width=6.9in]{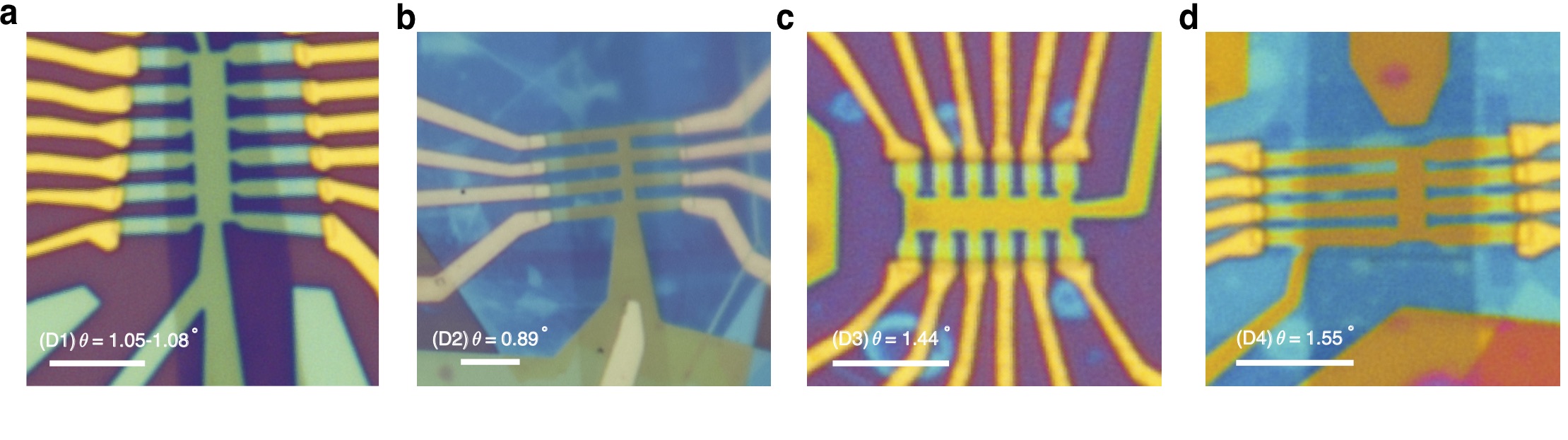} 
\caption{\textbf{Optical microscope images of the 4 tMBG devices.}
The twist angle of each device is denoted at the bottom left corner of each image. All scale bars are 5 $\mu$m.
}
\label{fig:S_devices}
\end{figure*}

\begin{figure*}[t]
\centering
\includegraphics[width=4.9in]{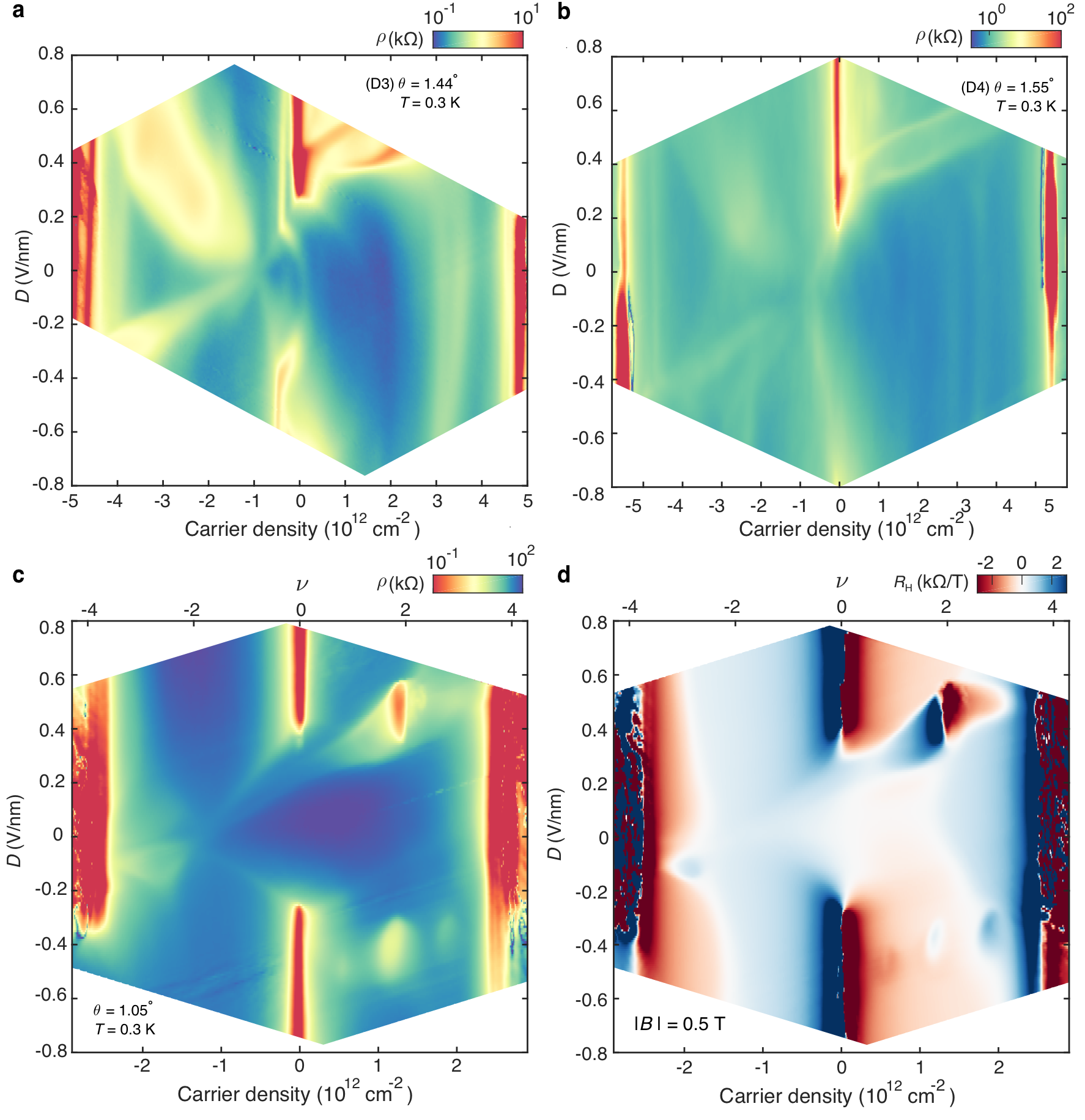} 
\caption{\textbf{Transport in tMBG at a variety of twist angles at $T=300$~mK.}
$\rho$ as a function of $n$ and $D$ for (\textbf{a}) device D3 ($\theta = 1.44^{\circ}$), and (\textbf{b}) device D4 ($\theta = 1.55^{\circ}$). At these twist angles, the gap at the charge neutrality points only opens for $D>0$ (pointing from monolayer to bilayer graphene). Features corresponding to single particle vHs can be seen in both conduction and valence bands. Although some features may arise owing to correlations, we do not observe any states with insulating behavior at integer $\nu$ within the bands.
\textbf{c-d} $\rho$ and $R_\mathrm{H}$ for device D1 using different contacts from Fig.~1d-e of the main text. The twist angle, $\theta = 1.05 ^{\circ}$, is slightly smaller than for the region of the device probed in the main text.
}
\label{fig:S_largerangle}
\end{figure*}

\begin{figure*}[t]
\includegraphics[width=\textwidth]{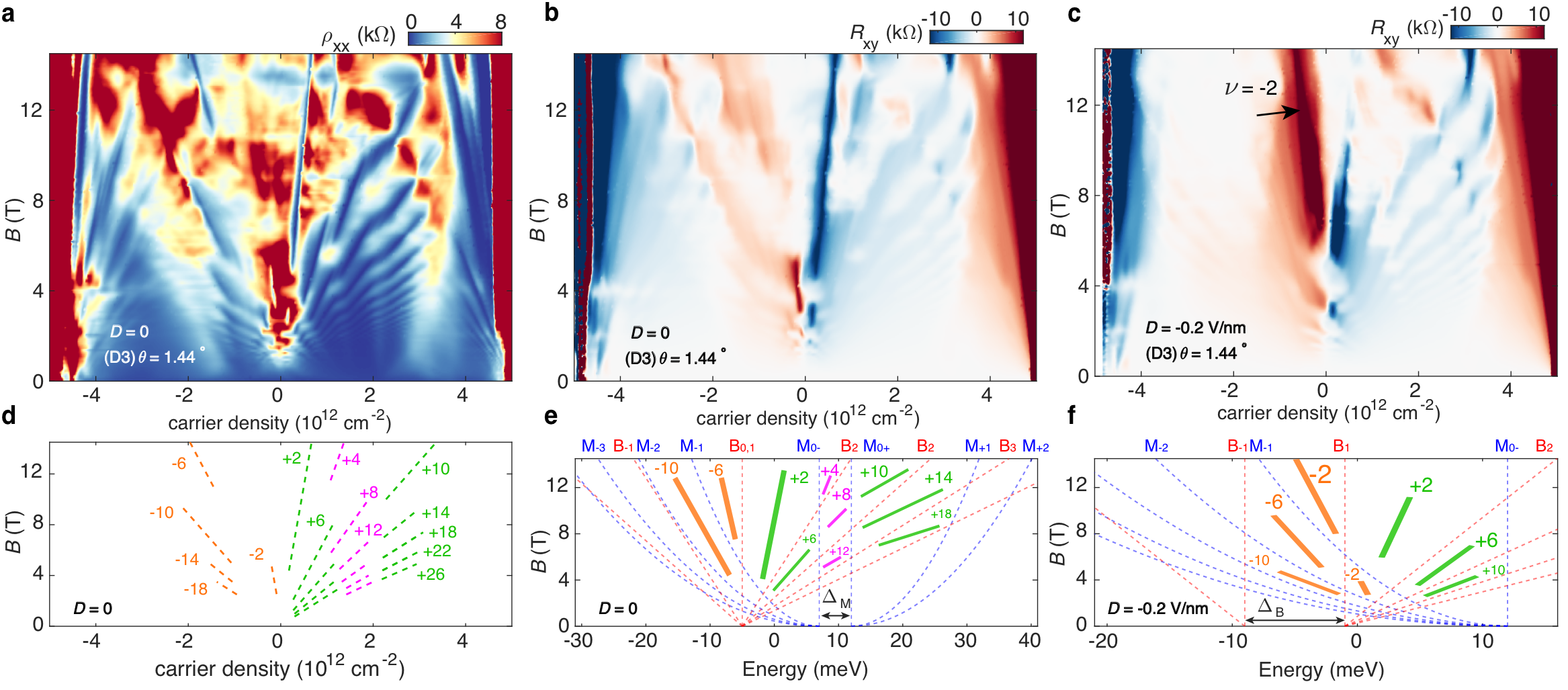} 
\caption{\textbf{Quantum oscillations in device D3 ($\theta = 1.44^{\circ}$).}
\textbf{a}, $\rho_{xx}$ and \textbf{b} $R_{xy}$ as a function of $n$ and $B_\perp$ at $D=0$.
\textbf{c}, Comparable $R_{xy}$ map at $D=-0.2$~V/nm.
\textbf{d}, Schematic illustration of the prominent quantum oscillations observed in \textbf{a}. Quantum oscillations exhibit four-fold degereracy at each value of $n$. For $n<0$, we observe a sequence of quantum oscillations with filling factors $\nu_{LL}=-2,-6,-10...$. For $n>0$, the sequence shifts from $+2,+6,+10...$ to $+4,+8,+12,...$ and back again as $n$ is raised. These sequences can be understood by considering the monolayer- and bilayer-like corners of the moir\'e Brillouin zone as approximately uncoupled owing to the larger twist angle, in which the dominant quantum oscillations are the sum of the contributions from each band.  
\textbf{e}, Dotted blue lines denote the Landau levels of the monolayer-like bands as a function of energy, following $E_{LL,m} = \sqrt{2e\hbar v_F^2 NB}$, where $N$ is the Landau level index and $v_F$ is the Fermi velocity. Dotted red lines denote the same for the bilayer-like bands, with $E_{LL,b} = \frac{e \hbar B}{m^*}\sqrt{N(N-1)}$, where $m^*$ is the effective mass. The total filling factor $\nu_{LL}$ within each gap is given by the sum of the Landau level indexes for each band (solid bars). The experimentally observed sequence of quantum oscillations is well reproduced taking $v_F = 1.44\times10^5$~m/s, $m^{*} = 0.14 m_0$, and including a charge neutrality band gap in the monolayer spectrum of $\Delta_M = 5$~meV and an offset between the monolayer and bilayer charge neutrality points of $\delta=14.5$~meV (indicative of band overlap).
\textbf{f}, Similar energy diagram for $D=-0.2$~V/nm. To account for the observation of $\nu_{LL}=-2$ at high magnetic field, a band gap for the bilayer spectrum of $\Delta_B=8$~meV is included, and we take $\delta=19.5$~meV.
}
\label{fig:S_1p44_Dall}
\end{figure*}

\begin{figure*}[t]
\includegraphics[width=4in]{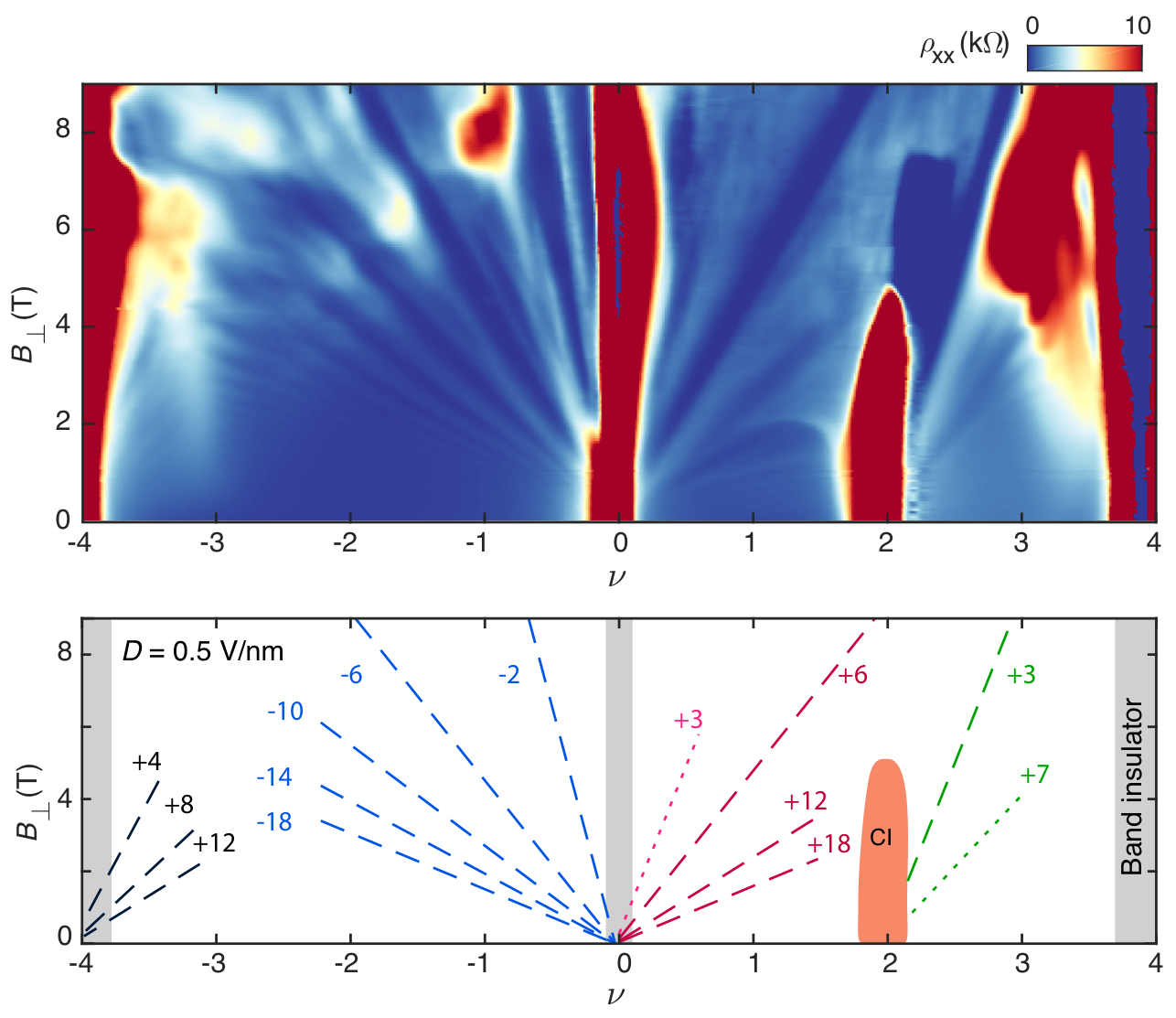} 
\caption{\textbf{Quantum oscillations at $D=+0.5$ V/nm in device D1 ($\theta=1.08^{\circ}$).}
Landau fan diagram up to full filling of the moir\'e unit cell with $D=+0.5$~V/nm at $T=300$~mK. Separate sequences of quantum oscillations emerge from $\nu=-4$ with a dominant sequence of $\nu_\mathrm{LL}=+4,+8,+12,...$, from $\nu=0$ with dominant sequences of $\nu_\mathrm{LL}=-2,-6,-10,...$ and $\nu_\mathrm{LL} = +6,+12,+18,...$, and from $\nu=+2$ with a dominant state at $\nu_\mathrm{LL} = +3$. Quantum oscillations emerging from $\nu=+2$ with larger filling factor do not follow a simple sequence, which may be a consequence of structural disorder in the sample. These states are illustrated schematically in the bottom panel with dashed lines. Dotted lines show additional states which emerge at higher field. Gray shaded regions denote band insulators and the orange shaded region denotes the CI state. The apparent 6-fold degeneracy of electron-like quantum oscillations emerging from $\nu=0$ may indicate the presence of multiple Fermi surface pockets within the moir\'e Brillouin zone.
}
\label{fig:DposQOs}
\end{figure*}

\begin{figure*}[t]
\includegraphics[width=4.65 in]{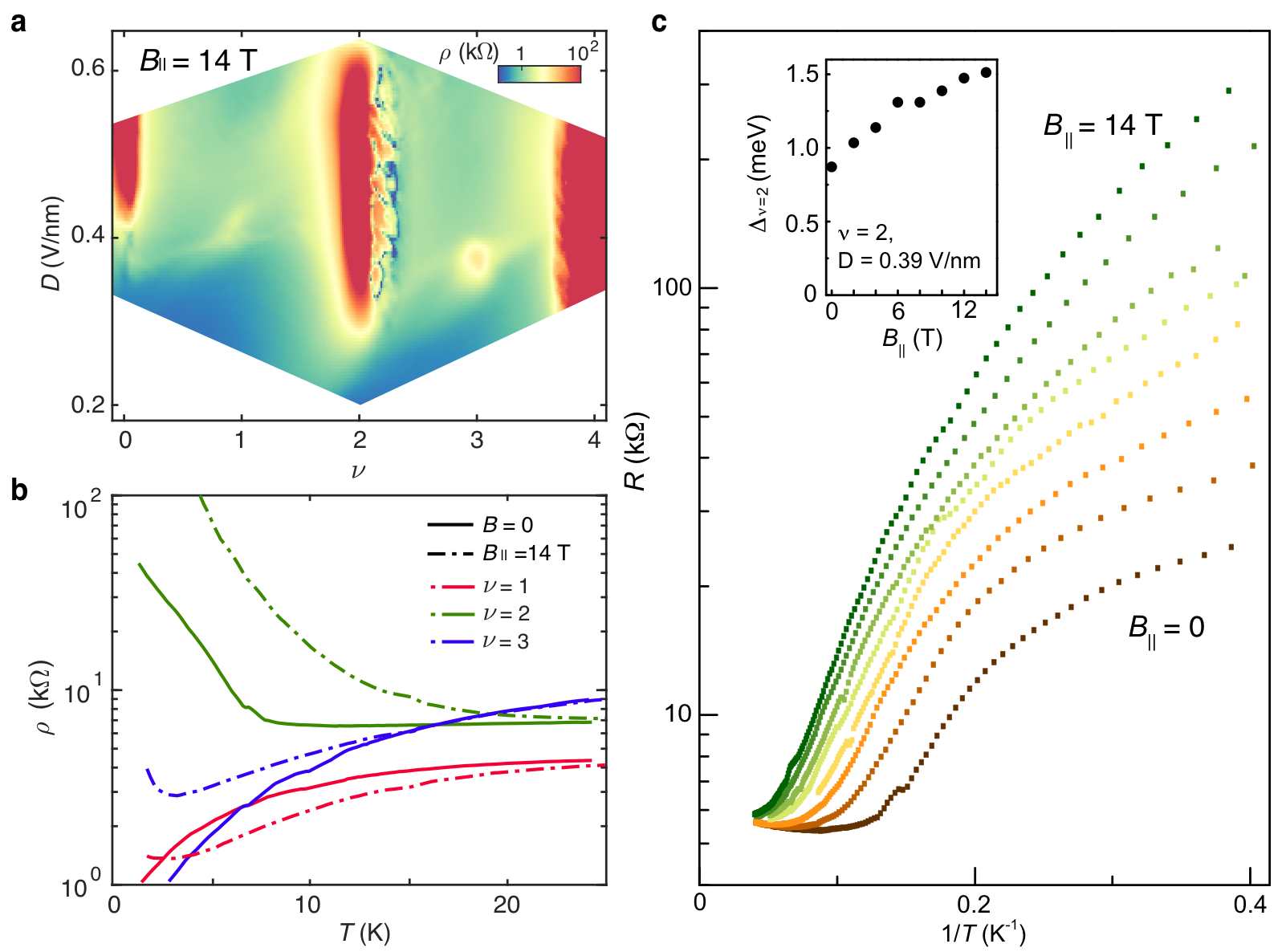} 
\caption{\textbf{Transport in an in-plane magnetic field in device D1.}
\textbf{a}, $\rho$ as a function of $\nu$ and $D$ at $B_{||}=14$~T.
\textbf{b}, $\rho(T)$ for $\nu=1,2,3$ at both $B=0$ (solid curves) and $B_{||}=14$~T (dashed curves). In the former, $\nu=2$ exhibits insulating behavior whereas $\nu=1$ and 3 are metallic. In the latter, all become more resistive at low temperature, and $\nu=1$ and 3 begin to undergo an insulating transition at low temperature.
\textbf{c}, Device resistance, $R$, versus $T^{-1}$ at $\nu=2$ and $D=+0.39$~V/nm. (Inset) Energy gaps, $\Delta_{\nu=2}$, as function of $B_{||}$, extracted from the thermal activation measurements. Gaps are extracted by fitting the data in the main panel to $R \propto e^{\Delta/2 k_B T}$, where $k_B$ is the Boltzmann constant. The gap grows with larger $B_{||}$ suggestive of a spin-polarized ground state. However, thermal activation measurements may be complicated by additional orbital contributions owing to the multilayer structure of tMBG~\cite{Lee2019,He2020}. 
}
\label{fig:S_parallelB}
\end{figure*}

\begin{figure*}[t]
\includegraphics[width=4.65 in]{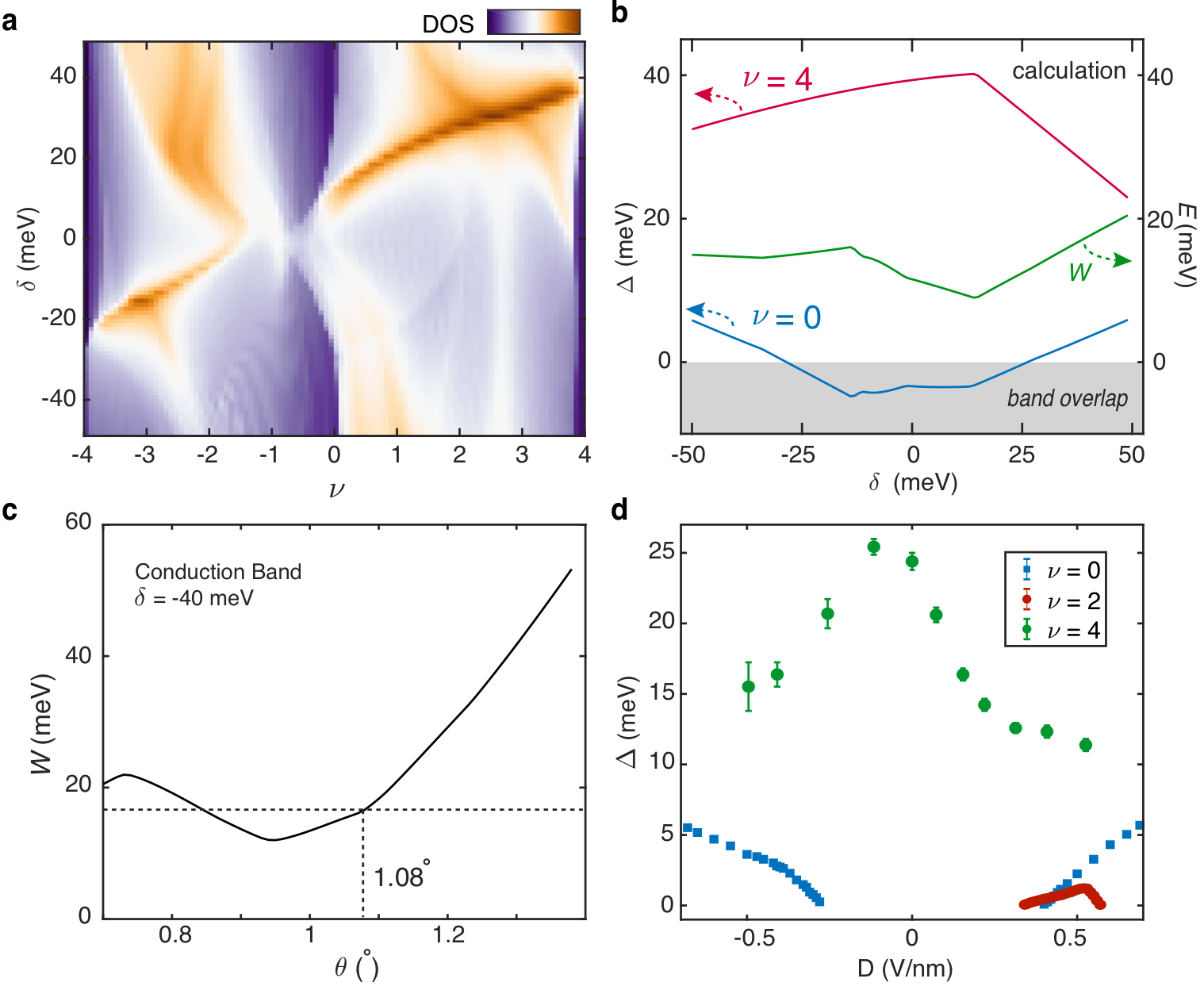} 
\caption{\textbf{Energy scales and density of states of tMBG.}
\textbf{a}, Single-particle density of states (DOS) calculated for $\theta = 1.08 ^{\circ}$ as a function of $\nu$ and $D$, following the model described in the text.
\textbf{b}, Calculated single particle gaps at $\nu=0,+4$ (left axis) as a function of interlayer potential, $\delta$. Bandwidth of the conduction band, $W$ (right axis). The bandwidth does not vary widely over our this range of $\delta$.
\textbf{c}, $W$ calculated as function of $\theta$ for $\delta=-40$ meV. $W$ does not change substantially for $0.85^{\circ} < \theta < 1.1^{\circ}$.
\textbf{d}, Experimentally measured energy gaps of $\nu=0,+2,+4$ as a function of $D$ in device D1 ($\theta = 1.08 ^{\circ}$). The single particle gaps at $\nu=0,+4$ are qualitatively consistent with the predictions in \textbf{b}. A CI gap at $\nu=+2$ only emerges over a finite range of $D>0$.
}
\label{fig:S_calculated_energy}
\end{figure*}

\begin{figure*}[t]
\includegraphics[width=3.5in]{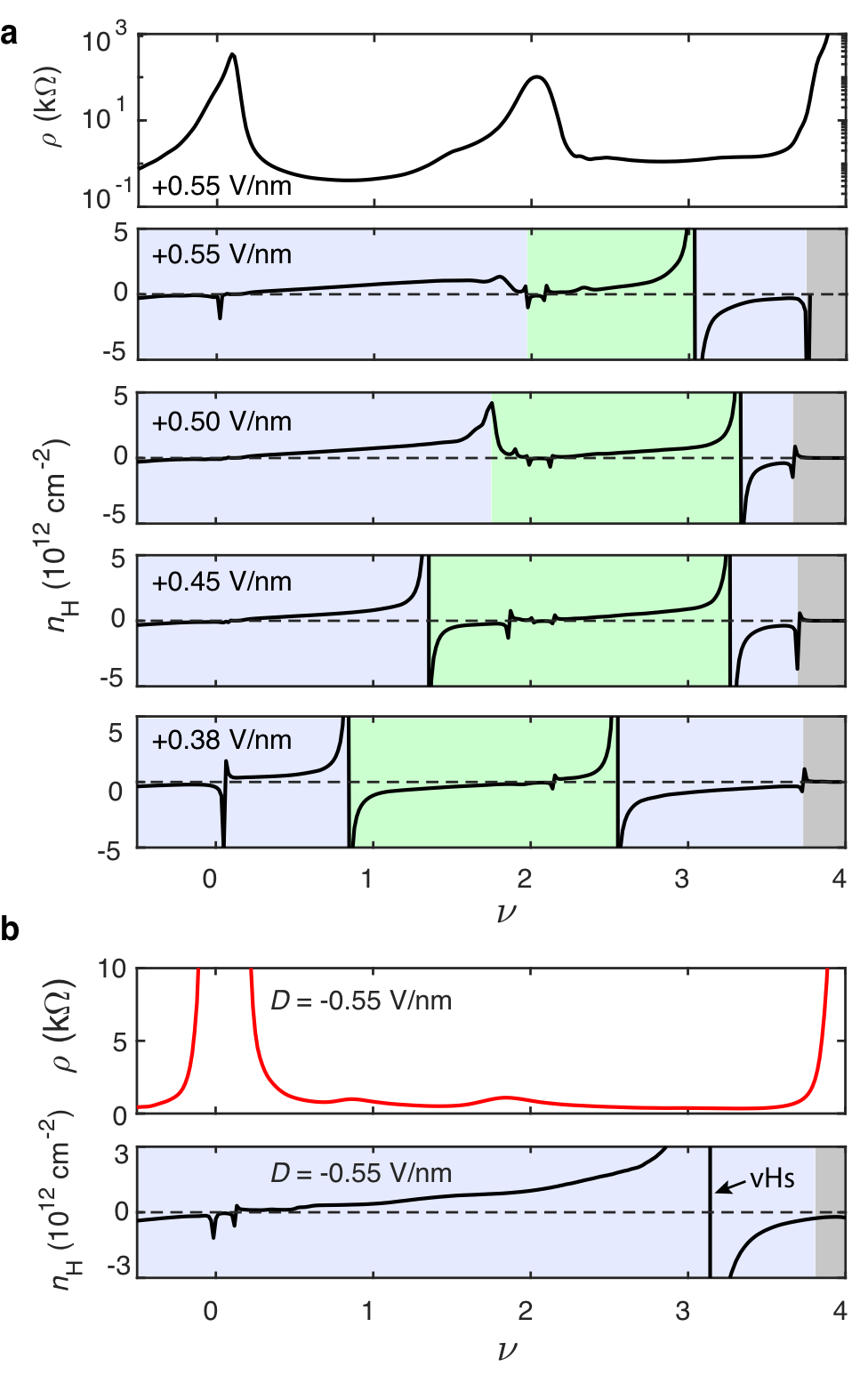} 
\caption{\textbf{Additional transport in device D1.}
\textbf{a}, $\rho$ at $D=+0.55$~V/nm (top panels) and $n_\mathrm{H}$ at various $D>0$ (bottom panels). Following the convention in the main text, green denotes regions in which spontaneous symmetry breaking renormalizes the Fermi surface.
\textbf{b}, $\rho$ (top) and $n_\mathrm{H}$ (bottom) at $D=-0.55$ V/nm. Signatures of correlated states are nearly absent at this $D$, and correspondingly $n_\mathrm{H}$ does not exhibit any abrupt sign changes or substantial deviations from a linear gate induced charge density, $n$. The sign change in $n_\mathrm{H}$ for $\nu$ just greater than 3 follows from a single particle model corresponding to a vHs, in which complete filling of a closed band requires an inversion of the sign of the charge carrier mass. 
}
\label{fig:S_HallDen}
\end{figure*}

\begin{figure*}[t]
\includegraphics[width=4.65 in]{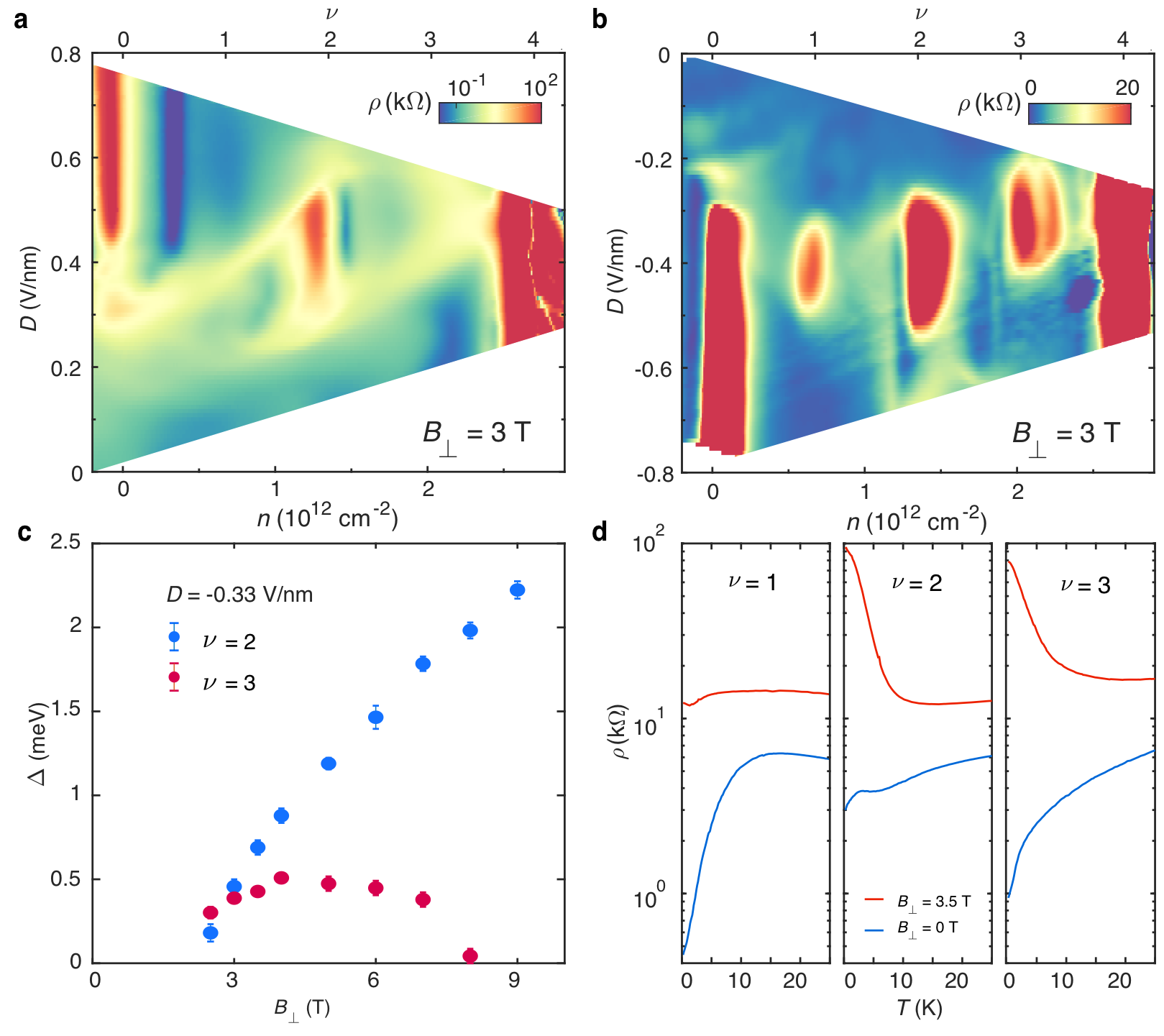} 
\caption{\textbf{High magnetic field transport in device D1.}
\textbf{a} $\rho$ as a function of $n$ and $D$ for $D>0$ at $B_\perp=3$~T. In addition to the CI state at $\nu=2$, vertical blue stripes correspond to the formation of Landau levels within the symmetry broken halo region. 
\textbf{b}, $\rho$ for $D<0$ at $B_\perp=3$~T. Field-assisted CI states emerge over a finite range of $D$, along with associated quantum oscillations.
\textbf{c}, Energy gaps of $\nu=2$ and 3 at $D=-0.33$~V/nm as a function of $B_\perp$, measured by thermal activation. The gap at $\nu=3$ closes at high field, which may be related to the spin and valley ordering of the state, and/or competition with Landau level formation.
\textbf{d}, $\rho(T)$ corresponding to $\nu=1,2,3$ at $B_\perp=0$ and 3.5~T. All exhibit unusual metallic temperature dependence at $B=0$, but are insulating ($\nu=2,3$) or near a crossover point $\nu=1$ at high field.
}
\label{fig:S_Bperp_all}
\end{figure*}

\begin{figure*}[t]
\includegraphics[width=5in]{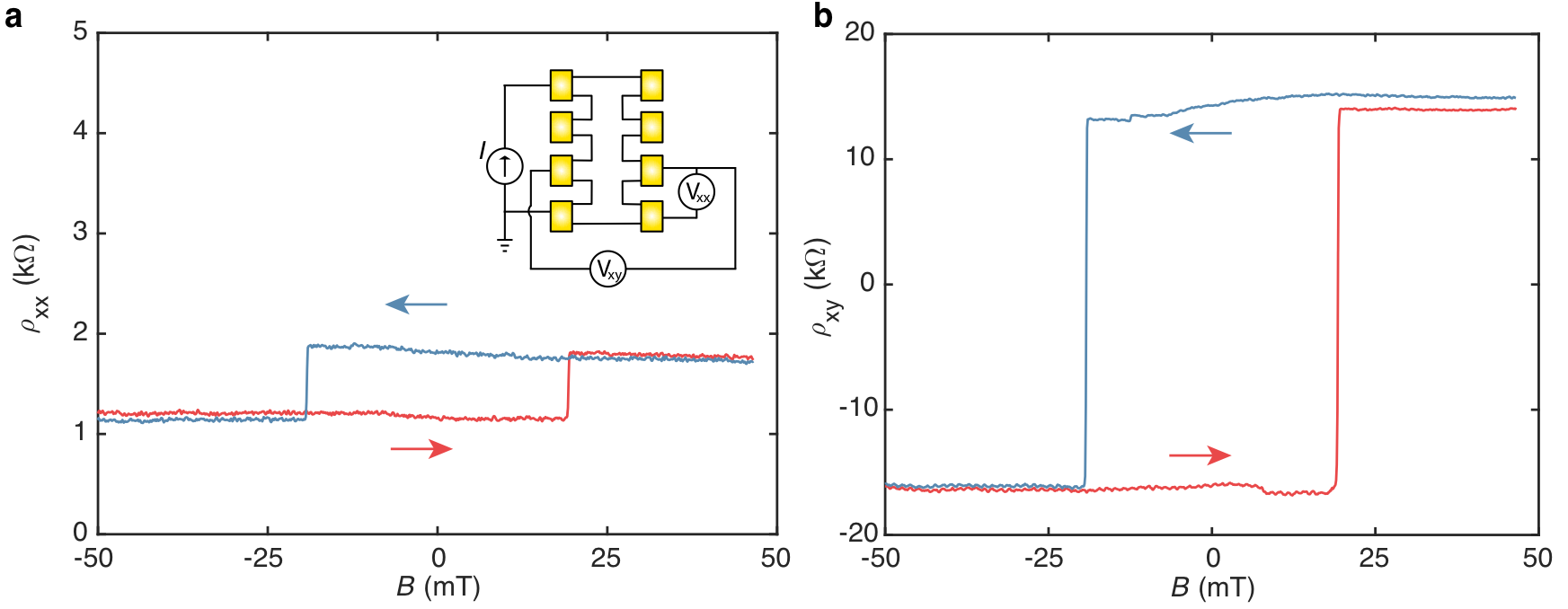} 
\caption{\textbf{Raw data of the AHE in device D2 ($\theta=0.89^{\circ}$).}
Data corresponds to Fig.~\ref{fig:4}b of the main text, however $\rho_{xx}$ is not symmetrized in \textbf{a}, and $\rho_{xy}$ not antisymmetrized in \textbf{b}. As a result, there is weak mixing between the two. The inset shows the measurement setup.
}
\label{fig:S_ahe_raw}
\end{figure*}

\begin{figure*}[t]
\centering
\includegraphics[width=7in]{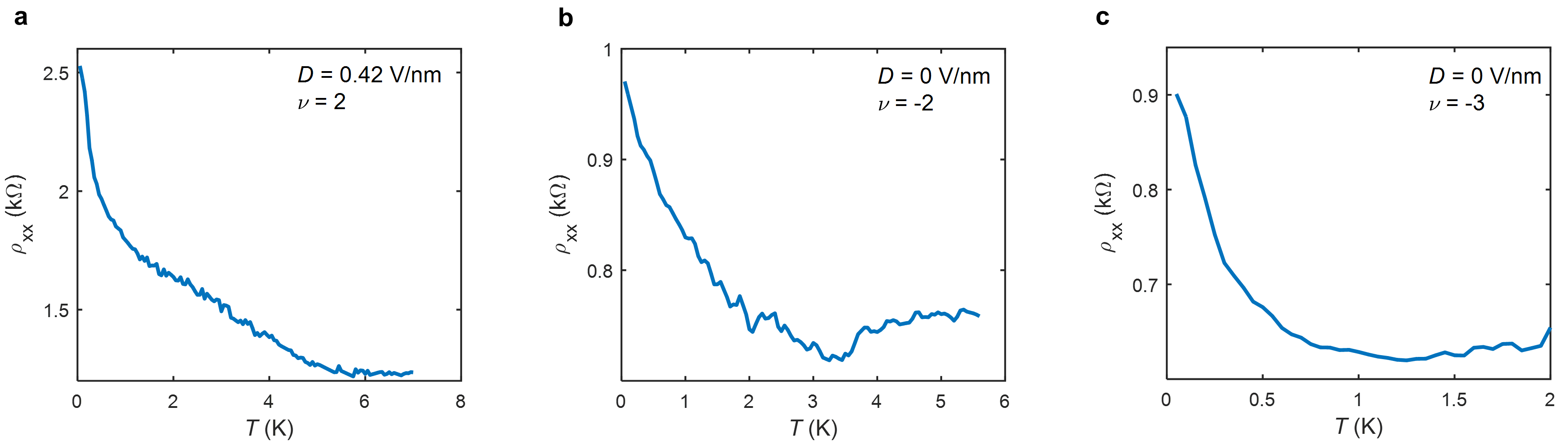} 
\caption{\textbf{Transport of the CI states at $\nu=+2,-2,-3$ in device D2.}
$\rho_{xx}(T)$ acquired at \textbf{a}, $\nu=2$ and $D$ = 0.42 V/nm, \textbf{b}, $\nu=-2$ and $D=0$, and \textbf{c}, $\nu=-3$ and $D=0$. All three states exhibit very weak insulating behavior at low $T$.
}
\label{fig:S_ahe_RvT}
\end{figure*}

\begin{figure*}[t]
\includegraphics[width=3in]{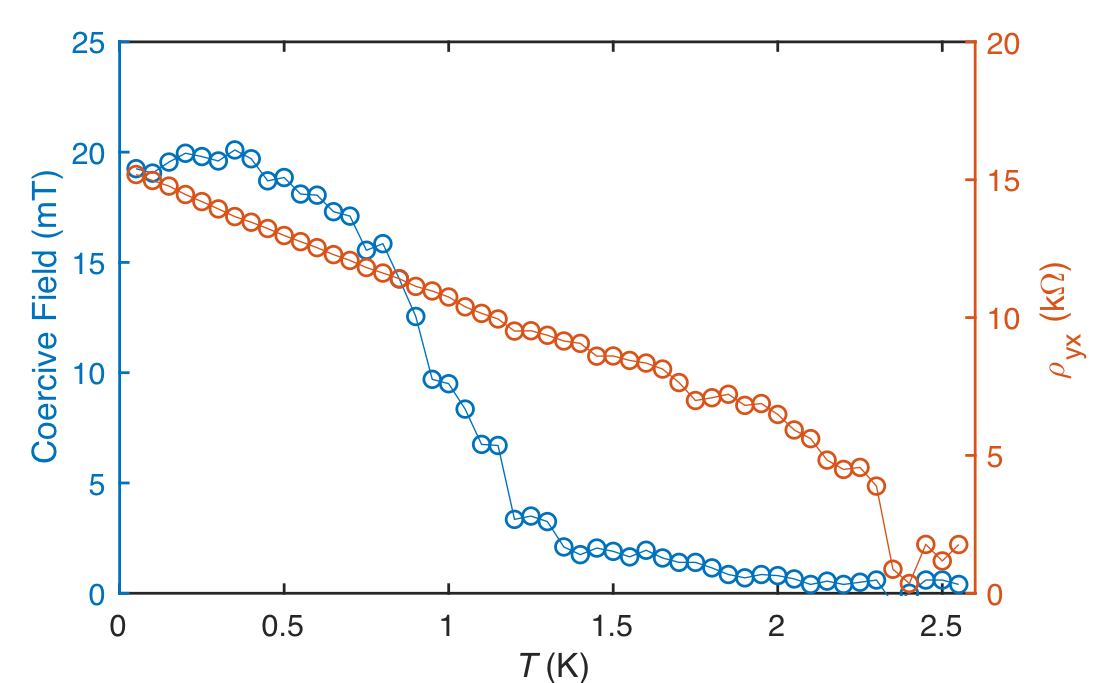} 
\caption{\textbf{Temperature dependence of the AHE resistivity and coercive field in device D2.} 
Measurements correspond to the data set shown in Fig.~\ref{fig:4}b of the main text, with $\nu=0.94$ and $D=+0.415$~V/nm.
}
\label{fig:S_ahe_Tdep}
\end{figure*}

\begin{figure*}[t]
\includegraphics[width=3in]{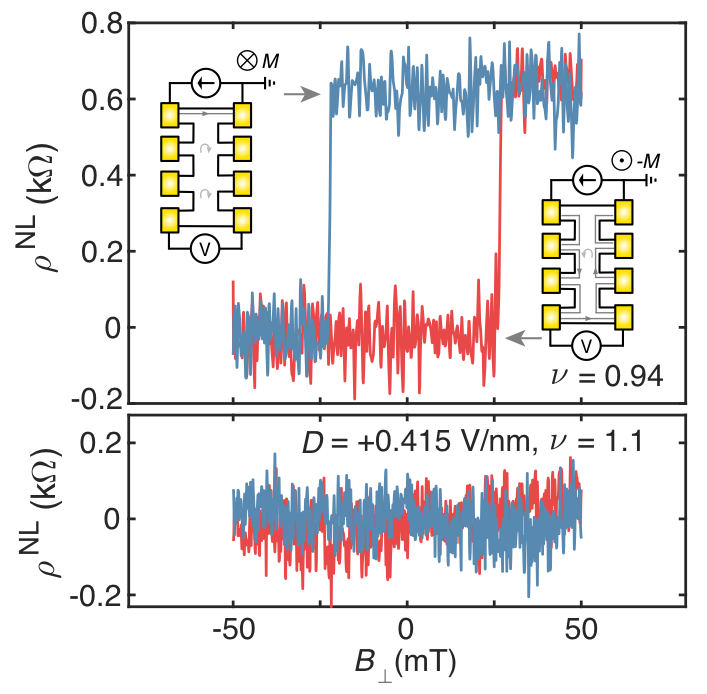} 
\caption{\textbf{Non-local resistance measurements in device D2.}
Non-local resistance, $\rho^{NL}$, acquired at $D=+0.415$~V/nm and $\nu=0.94$ (top) and $\nu=1.1$ (bottom). Schematics illustrate the contact configuration used for the measurement, as well as the corresponding chiral edge modes equilibrated to the source electrode. Outside the magnetic regime (bottom panel), vanishingly small $\rho^{NL}$ is observed. In contrast, substantial $\rho^{NL}$ is observed for a single sign of the magnetization. This has previously been understood to result from differences in equilibration between the nonlocal voltage probes depending on the chirality of the edge modes~\cite{Kou2014,Sharpe2019}, indicating the existent of these topological edge modes in our device.
}
\label{fig:S_ahe_NL}
\end{figure*}

\begin{figure*}[t]
\includegraphics[width=4.5in]{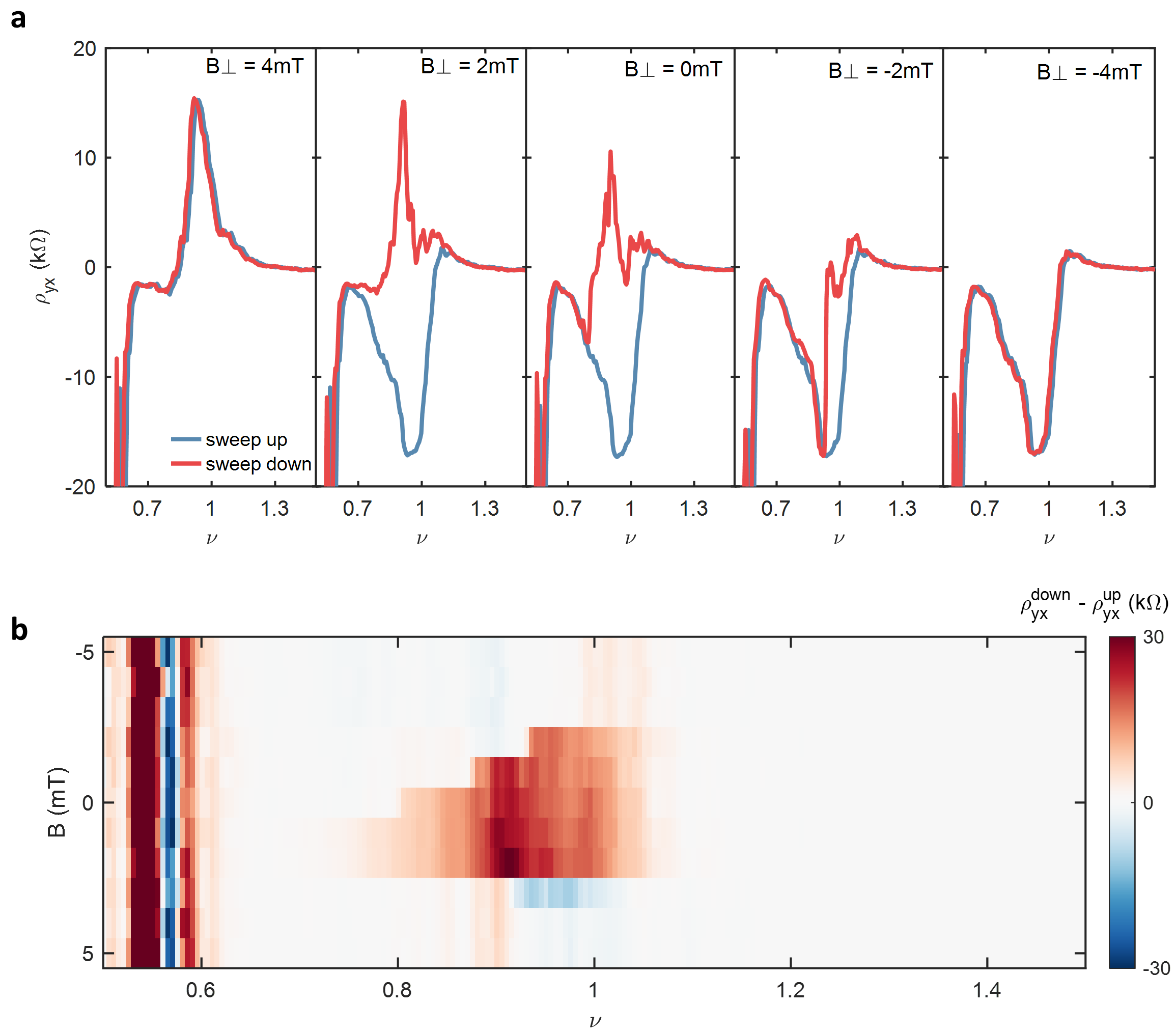} 
\caption{\textbf{Doping-induced switching of the magnetic order with $B_\perp$ in device D2.}
\textbf{a}, $\rho_{yx}$ as $\nu$ is swept back and forth from -0.5 to +2.5 at $D=+0.415$~V/nm, with $B_{\perp}$=$+4$~mT, $+2$mT, 0mT, $-2$~mT, $-4$~mT. Doping-induced switching of the magnetic order is only observed within a small range of $B_{\perp}$ from $\sim +2$~mT to $-2$~mT. At larger $B_\perp$, the magnetic state is aligned with the applied field regardless of the sweep direction.
\textbf{b}, The difference between $\rho_{yx}$ sweeping up and down as a function of $B_{\perp}$. The red region denotes the regime in which the magnetization can be switched with doping.
}
\label{fig:S_ahe_InitiWithB}
\end{figure*}

\begin{figure*}[t]
\centering
\includegraphics[width=7in]{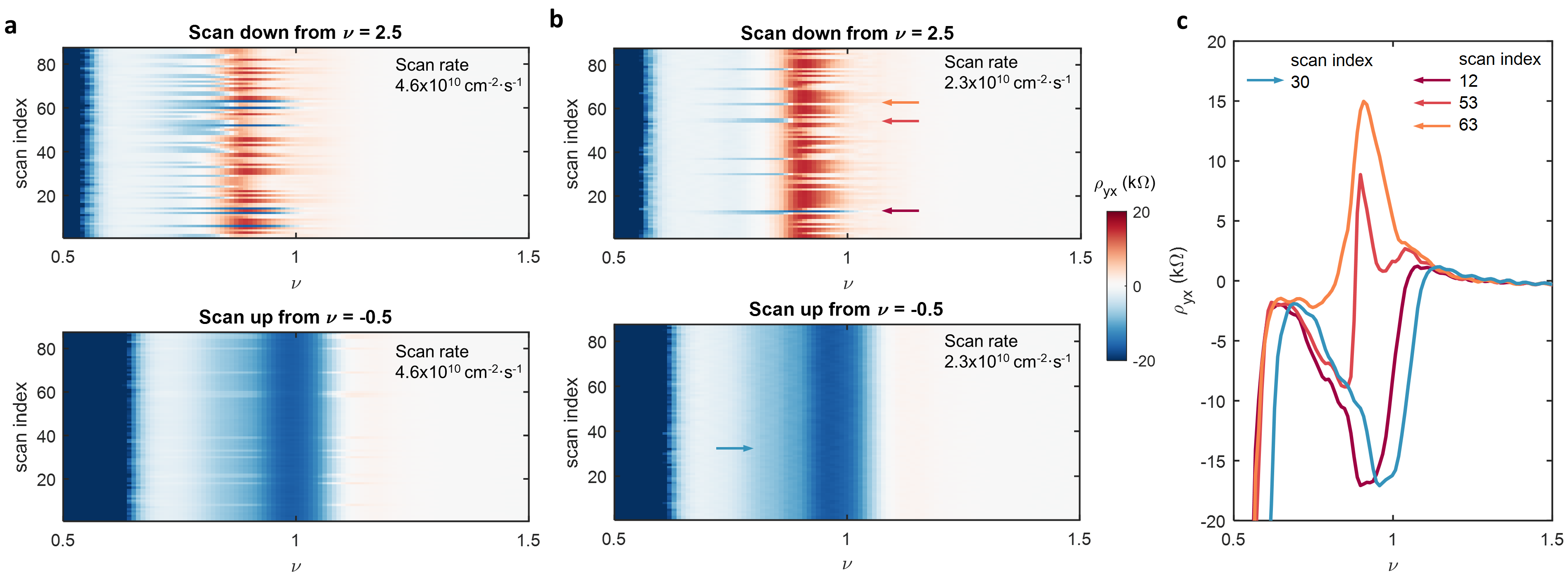} 
\caption{\textbf{Stability of the doping-induced switching of the magnetic order in device D2.}
\textbf{a}, $\rho_{yx}$ as $\nu$ is swept down from $\nu=2.5$ (top) and up from $\nu=-0.5$ (bottom). The sweep is repeated 87 times, denoted by the scan index. The offset of the features in $\nu$ between the two maps arise owing to fast sweeping of $\nu$, as compared with the integration time of the lock-in amplifier. In the sweeps down from $\nu=2.5$, the magnetic state is not successfully switched to the positive state (red) every time. 
\textbf{b}, Comparable measurement but with $\nu$ swept at half the rate as in \textbf{a}. The switching to the positive state (red) occurs in all but one sweep with this reduced sweeping rate.
\textbf{c}, Selected line cuts corresponding to the arrows in \textbf{b}, showing various representative behaviors of $\rho_{yx}$ as $\nu$ is swept. When sweeping up from $\nu=-0.5$, the magnetic state is negative every time, and closely resembles the blue curve (scan index 30). A wider range of behaviors is observed upon sweeping down from $\nu=2.5$. Most frequently, the magnetic state closely resembles scan index 63, in which the positive state is observed. On occassion, the state appears to switch through various metastable configurations (scan index 53), corresponding to regions with trailing blue streaks in the top panel of \textbf{b}. Finally, in 1 of the 87 traces, we did not observe switching at all (scan index 12).
}
\label{fig:S_ahe_stable}
\end{figure*}

\begin{figure*}[t]
\centering
\includegraphics[width=6.5in]{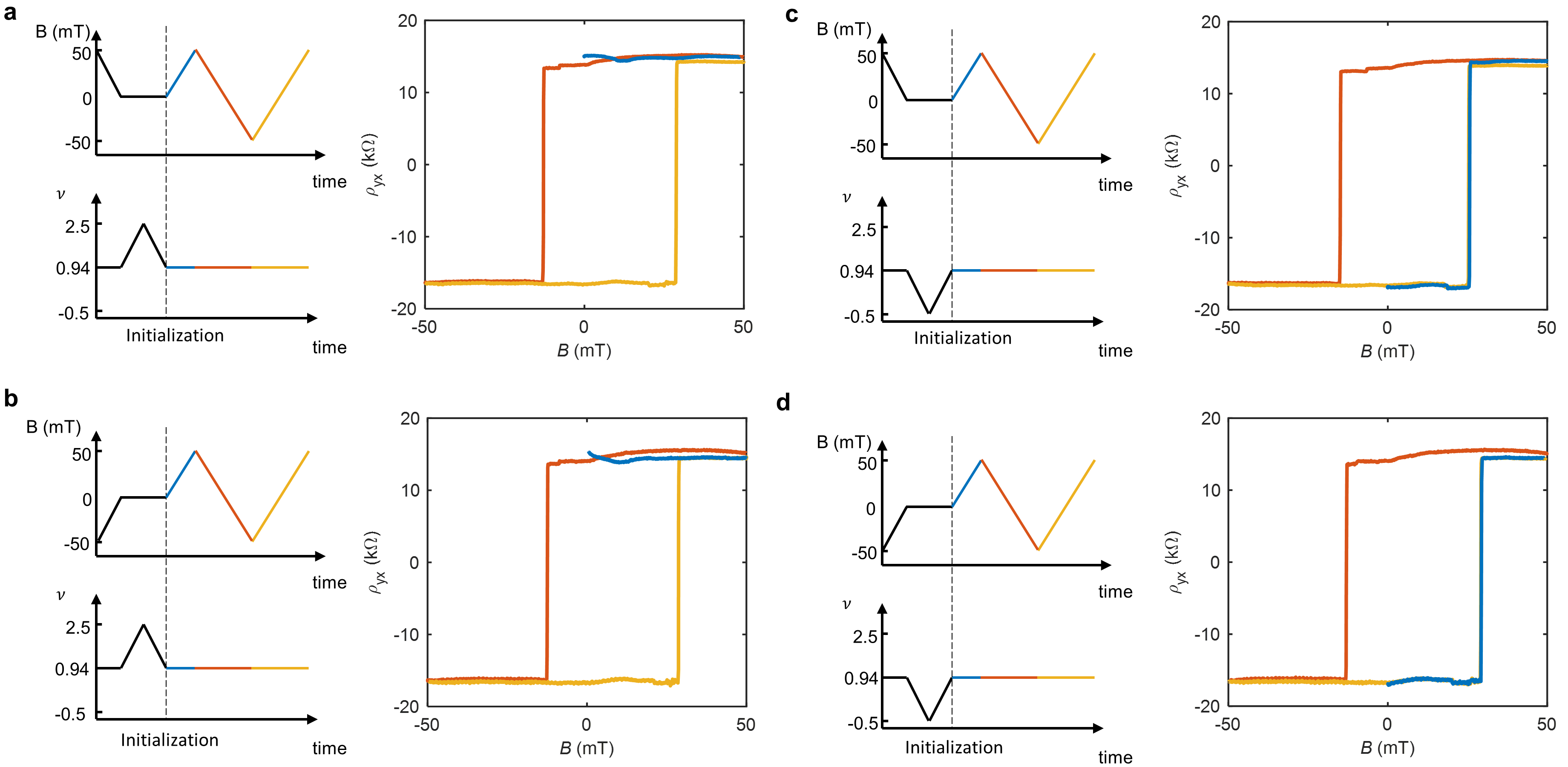} 
\caption{\textbf{Dependence of doping-induced switching on magnetic field initialization in device D2.}
\textbf{a-d}, show hysteresis loops at $\nu=0.94$ and $D=+0.415$~V/nm, acquired after initiating the state with different magnetic field and doping trainings. The schematics to the left of each $\rho_{yx}$ panel denote the sequence of $B_\perp$ and $\nu$ sweeps as a function of time. Black lines denote the initialization operations, performed prior to data acquisition. The initialized magnetic state is indicated by the starting point of the blue $\rho_{yx}$ curve at $B=0$. Subsequent to the initialization, $\nu$ is held fixed at 0.94 while $B_\perp$ is swept back and forth. Notably, independent of the field training history, $\rho_{yx}$ is initialized to the positive state when $\nu$ is swept down from 2.5, and to the negative state when $\nu$ is swept up from $-0.5$. Subsequent magnetic field sweeping forms a similar hysteresis loop independent of the initialized state. This helps to rule out any effects of the magnetic field initialization in our observations of the doping-induced switching effect at $B=0$.
}
\label{fig:S_ahe_fieldscans}
\end{figure*}

\begin{figure*}[t]
\centering
\includegraphics[width=6in]{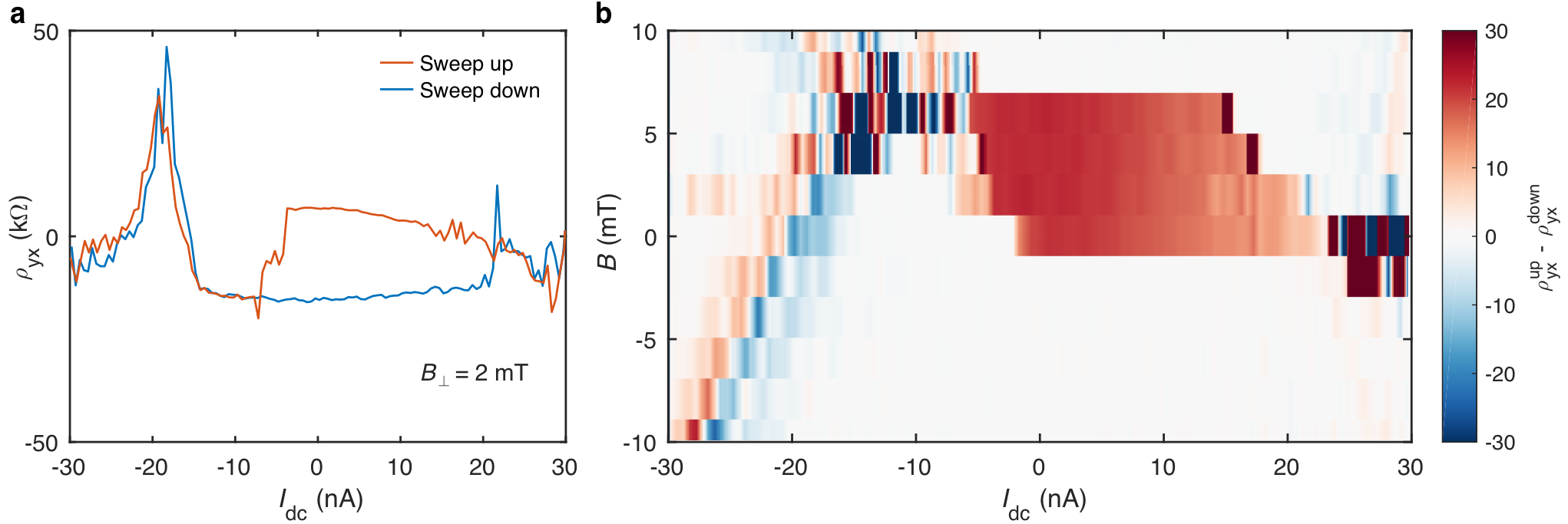} 
\caption{\textbf{Switching the magnetic ordering with a direct current bias in device D2.}
\textbf{a}, $\rho_{yx}$ as the dc current, $I_{dc}$, is swept back and forth at $B_\perp=2$~mT. The magnetic order can be switched with $I_{dc}$. Additional resistive states appear at larger $I_{dc}$ and may be related to current melting of the magnetic ordering, however these features are not currently well understood.
\textbf{b}, The difference between $\rho_{yx}$ sweeping up and down as a function of $B_\perp$. The red region denotes the regime in which the magnetization can be switched with $I_{dc}$. At high field, the magnetization aligns with the external field and cannot be switched with $I_{dc}$.
}
\label{fig:S_ahe_dcswitch}
\end{figure*}

\end{document}